 \documentclass[final,5p,times]{elsarticle}

\usepackage{amssymb}

\usepackage{lineno}

\usepackage{graphicx}
\usepackage{subcaption}
\usepackage{float}
\usepackage[thinc]{esdiff}
\usepackage{amsmath}
\usepackage{nicefrac}
\usepackage{mathtools}
\usepackage{appendix}
\usepackage{dcolumn}
\usepackage{bm}
\usepackage{widetext}

\begin{document}

\begin{frontmatter}



\title{The Economics of Interstellar Flight}


\author{Philip Lubin}
\ead{lubin@ucsb.edu}

\author{Alexander N. Cohen}
\affiliation{organization={Department of Physics, University of California - Santa Barbara},
            addressline={Broida Hall}, 
            city={Santa Barbara},
            postcode={93106}, 
            state={California},
            country={United States}}

\begin{abstract}
Large scale directed energy offers the possibility of radical transformation in a variety of areas, including the ability to achieve relativistic flight that will enable the first interstellar missions, as well as rapid interplanetary transit. In addition, the same technology will allow for long-range beamed power for ion, ablation, and thermal engines, as well as long-range recharging of distant spacecraft, long-range and ultra high bandwidth laser communications, and many additional applications that include remote composition analysis, manipulation of asteroids, and full planetary defense. Directed energy relies on photonics which, like electronics, is an exponentially expanding growth area driven by diverse economic interests that allows transformational advances in space exploration and capability. We have made enormous technological progress in the last few years to enable this long-term vision. In addition to the technological challenges, we must face the economic challenges to bring the vision to reality. The path ahead requires a fundamental change in the system designs to allow for the radical cost reductions required. To afford the full-scale realization of this vision we will need to bring to fore integrated photonics and mass production as a path forward. Fortunately, integrated photonics is a technology driven by vast consumer need for high speed data delivery. We outline the fundamental physics that drive the economics and derive an analytic cost model that allows us to logically plan the path ahead. 
\end{abstract}




\end{frontmatter}


\section{\label{sec:intro}Introduction}

    One of the dreams of humanity is to travel to the stars and begin visiting the many exoplanets within our reach. With the number of planets per star being approximately unity based on the latest Kepler data and with even our nearest stellar neighbor, the Alpha Centauri system, having at least one confirmed exoplanet, the possibility of reaching interstellar targets is a dream we can begin to seriously explore. As we have outlined in a series of papers, the ability to travel to and explore nearby stellar systems requires a radical change in both propulsion systems and in spacecraft design. The ability to achieve the speeds required is becoming a possibility due to recent advances in directed energy (DE) systems that allow us to remove the propulsion system and its associated mass from the spacecraft. The transformations that will come from this approach allow a radical change in capability. It is critical to understand that this program has a series of ``steps'' with increasing capabilities in speed and spacecraft mass and that the program by its nature is both revolutionary and evolutionary with no end point. We will see there is a trade space where the fundamental physics of the problem drive economic tradeoffs. Given the exponential growth in some of the technologies but not in others and thus the dynamic nature of the system costs, we rapidly conclude that the ``time of entry'' to achieve a given milestone or ``step along the ladder'' is critical to understand. Much like the semiconductor industry, we need to develop an ``interstellar roadmap'' that combines a photonics and electronics roadmap with a detailed understanding of the non-exponential aspects of the program. 

\subsection{Why Directed Energy Propulsion is Needed}
    
\begin{figure}
        \centering
        \includegraphics[width=0.45\textwidth]{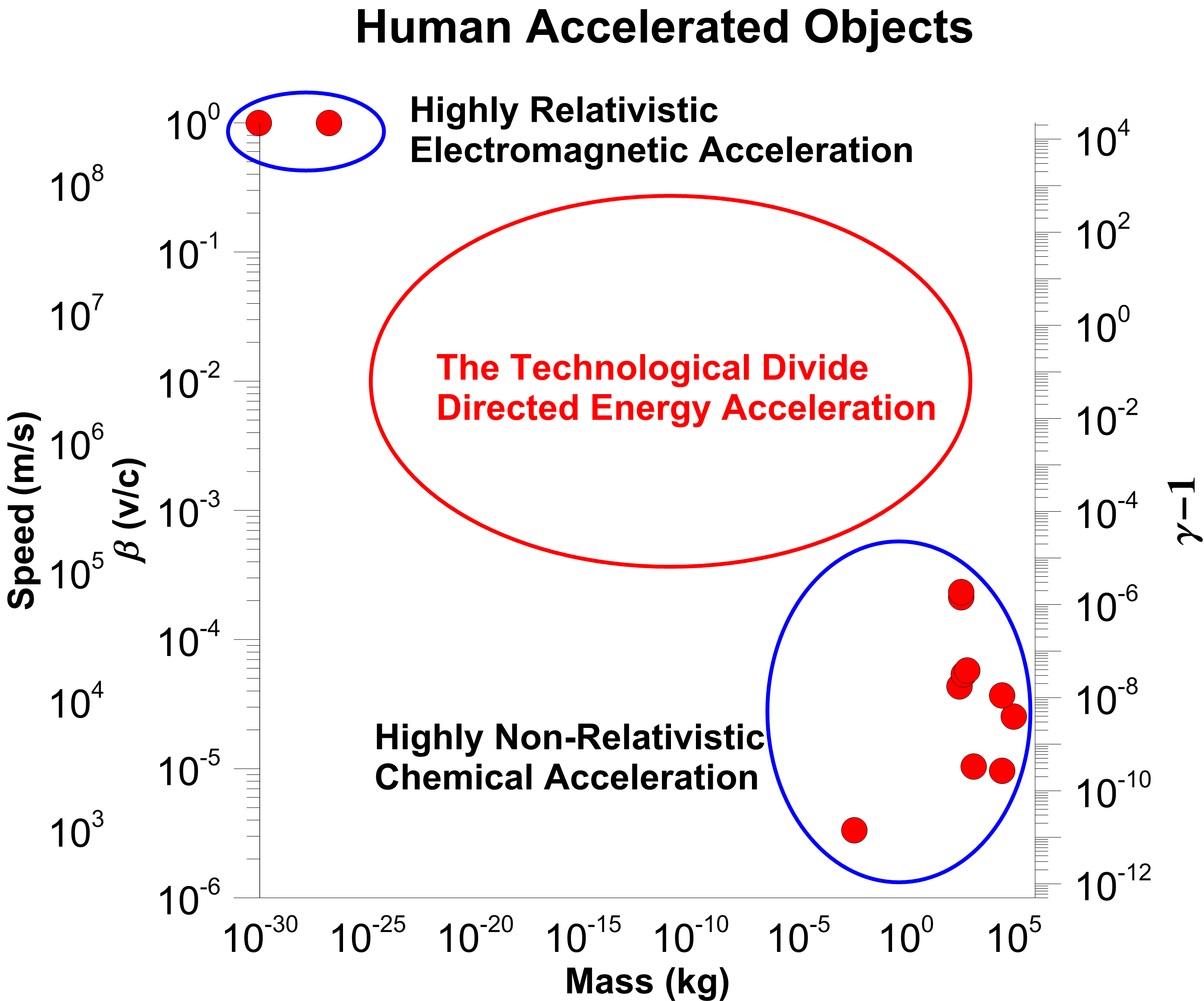}
        \caption{Speed and fractional speed of light achieved by human accelerated objects vs. mass of object from sub-atomic to large macroscopic objects. Right side $y$-axis shows $\gamma-1$ where $\gamma$ is the relativistic ``gamma factor.'' $\gamma-1$ times the rest mass energy is the kinetic energy of the object.}
        \label{fig:humanacceleratedobjects}
    \end{figure}

    Chemical energy released through the formation and breaking of chemical bonds forms the basis for every space launch that has left the Earth. The energy available in chemical bonds is very small compared to the relativistic rest mass energy of the chemical used. Typically, each chemical bond, whether in food or for planes or rockets, is only about 1eV per bond or typically one ten billionth of the rest mass of the constituent chemical. The implications of this for achieving high speeds are so severe that literally, if we took the entire mass of the universe and converted it into chemical propellant or ``rocket fuel'' then the highest speed we could achieve for a payload that was a single proton, is about 170 times the exhaust speed. Typical chemical rocket engines achieve an exhaust speed of 2-4 km/s giving a final speed for our ``proton payload'' of about 300-600 km/s or about 0.1\% the speed of light. Again, this is for a single proton payload being driven with the entire universe as ``rocket fuel.'' This is nowhere close to being relativistic. To reach relativistic speeds we would need GeV per bond equivalent or about a billion times larger than chemical reactions. Directed energy, being light, is only limited by the speed of light and hence to reach speeds approaching the speed of light we cannot use chemistry based propulsion. This difference in speeds achieved is dramatically illustrated if we compare the beta ($v/c$) and gamma factors achieved (Figure \ref{fig:humanacceleratedobjects}). We clearly have the ability to produce highly relativistic systems using particle accelerators, but only at the particle level. Practical systems need to be macroscopic and we do not currently have the technological means to self-assemble relativistic particles into macroscopic systems. A more complex analysis yields the conclusion that even nuclear fission and fusion based engines cannot achieve relativistic speeds. Only two candidates consistent with known physics allow relativistic flight. One is the possible use of matter-antimatter annihilation engines and the other is directed energy. We do not have a current path to annihilation engines. We do have a path forward with directed energy \citep{Lubin2015,Lubin2016c,Lubin2018,Lubin2019,Bible2013,Hughes2014,Hettel2019,Brashears2015a,Srinivasan2016}.
    
    {}
    
\subsection{Finite Milestones in an Infinite Journey}

    Unlike many programs, this program has no terminal point. The ability to make and deploy increasingly capable DE systems in both power and aperture size means there is no stopping point in the future, but simply a series of increasingly ambitious systems that enable increasingly ambitious and capable missions enabled by this transformation. Because this is an infinite program we need to look at each step along the path as both a ``rung'' in an infinite ladder as well as a well-defined milestone with specific capabilities. Each step will be different, building on the lessons learned from the previous step and each step will incorporate increasingly sophisticated technology as well as decreasing costs per unit of measure. Thus, it becomes critical to begin a detailed economic planning phase that is flexible enough to allow transformative shifts that may occur as we proceed, but is cognizant of the areas where such disruptive technology coupled with economic advances are required for a given ``step of the ladder.'' It is this infinite ladder combined with exponential growth which represents a unique opportunity as well as planning challenge. Start ``too early'' and it costs too much, start ``too late'' and you loose mission opportunity. Hence the need for careful planning of technology, economics, and mission desires.
    
\section{Directed Energy Approaches}

\begin{figure}[H]
        \centering
        \includegraphics[width=0.45\textwidth]{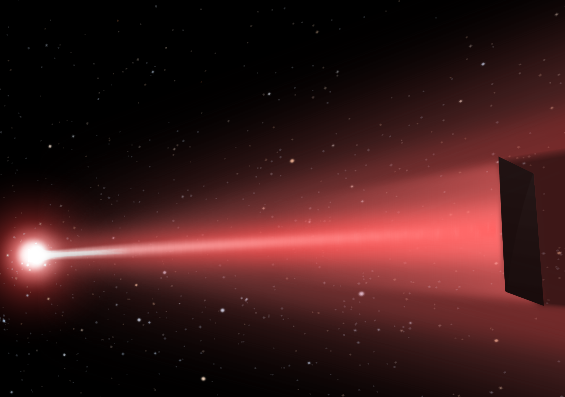}
        \caption{Artistic rendering of a laser-driven reflector. Credit: Q. Zhang -- UC Santa Barbara.}
        \label{fig:artisticrendering}
    \end{figure}

    Using directed energy for propulsion opens up two fundamental types of missions spaces. Both types use the same basic DE technological infrastructure and one DE system can be used for both modes. The economic analyses are similar, but not identical. This paper concentrates on the first mode below:
    \begin{itemize}
        \item[-] Direct Drive Mode (DDM): DE is used in a ``momentum transfer mode'' for propulsion via direct momentum transfer to the spacecraft via a reflective sail. This is the approach required for relativistic and hence interstellar flight, and is illustrated in Figure \ref{fig:artisticrendering}.
        \item[-] Indirect Drive Mode (IDM): DE is used in an ``energy transfer mode'' for propulsion via energy conversion to electrical energy onboard the spacecraft where the electrical energy is used to drive an ion engine that then provides the thrust. This mode requires ``fuel'' to be expelled (mass ejection) and hence has limits on total momentum and is not suitable for relativistic flight. The IDM mode is most useful for high mass missions inside our solar system.
    \end{itemize}
    While the DDM approach has no upper payload mass limit, the IDM mode is generally more practical for missions that do not require extreme speeds as the thrust per unit power is $2/I_{sp}$ where the $I_{sp}$ is the specific impulse of the engine. For DDM approach the $I_{sp}$ is 60 million while for the IDM approach the $I_{sp}$ for ion engines is 1000 (Xe, etc.) to 50,000 (Li). This means that for the same mission thrust desired, an IDM approach uses much lower power BUT achieves much lower final speed. For solar system missions with high mass, the final speeds are typically of order 100 km/s and hence an IDM approach is generally economically preferred. Another way to think of this is that a system designed for a low mass relativistic mission can also be used in an IDM approach for a high mass, low speed mission. 
    
    {}
    
\section{Directed Energy Phased Array Laser (DEPA)}

    The key to this program is the ability to build a sufficiently powerful laser driver with a large enough effective aperture to allow the beam to stay on the spacecraft long enough to propel it to high speed. For relativistic flight ($>0.1 c$) development of ultra-low mass probes is also needed. Recent developments now make both of these possible. The photon driver is a laser phased array which eliminates the need to develop one extremely large laser and replaces it with a large number of modest laser amplifiers in a MOPA (Master Oscillator Power Amplifier) configuration with a baseline of Yb amplifiers operating at 1064 nm. The system is phase locked using a coherent beacon that is either carried by the spacecraft or reflects off the spacecraft. Maintaining phase integrity is one of the key challenges. This approach is analogous to building a super computer from a large number of modest processors. This approach also eliminates the conventional optics and replaces it with a phased array of small low cost optical elements. As an example, on the eventual upper end, a full scale system (50-100 GW) will propel a wafer-scale spacecraft with a meter class reflector (laser sail) to about $c/4$ in a few minutes of laser illumination allowing hundreds of launches per day or $10^5$ missions per year. Such a system would reach the distance to Mars (1 AU) in 30 minutes, pass Voyager I in less than 3 days, pass 1,000 AU in 12 days, and reach Alpha Centauri in about 20 years. The same system can also propel a 100 kg payload to about 0.01$c$ and a 10,000 kg payload to more than 1,000 km/s, though there is a practical trade space of using DDM vs IDM for larger mass spacecraft. These systems are vastly faster than any currently imagined conventional propulsion system including ion engines, solar sails, e-sails etc.
    
    The basic system topology is scalable to any level of power and array size where the tradeoff is between the spacecraft mass and speed and hence the ``steps on the ladder.'' One of the advantages of this approach is that once a laser driver is constructed it can be used on a wide variety of missions, from large mass interplanetary to low mass interstellar probes, and can be amortized over a very large range of missions. Additional tasks beyond photon propulsion include beamed power for ion engine systems, distant spacecraft recharging (eliminating RTG's in some cases), full standoff planetary defense against both asteroids and comets, solar system wide active illumination or laser scanning (LIDAR) to find and study smaller objects, asteroid manipulation and composition analysis, terraforming applications, and a path to extremely large telescopes. Many of these applications are discussed in detail in our many papers.
    
    \begin{figure*}
        \centering
        \begin{tabular}{c c c}
             \includegraphics[width=0.5\textwidth]{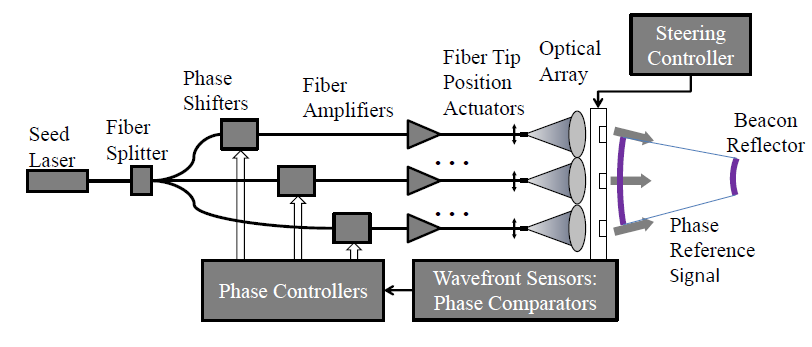} & \includegraphics[width=0.15\textwidth]{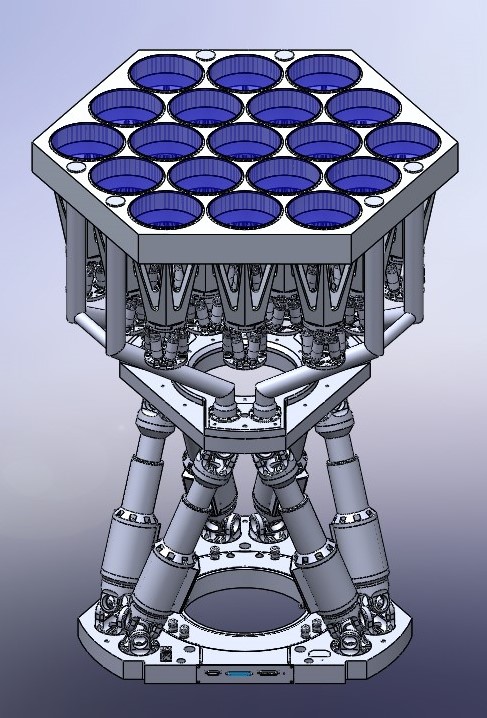} & \includegraphics[width=0.245\textwidth]{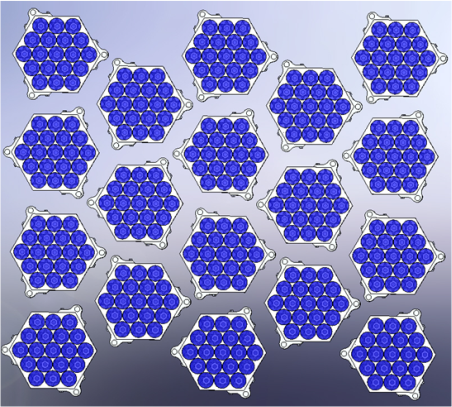}  \\
             (a) & (b) & (c) 
        \end{tabular}{}
        \caption{(a) Schematic design of phased array laser driver. Wavefront (phase) sensing from both local and extended beacons combined with the system metrology are critical to phasing (forming) the final diffraction limited beam. (b) Design of a one meter hexagonal module with 19 optical elements, each of which is connected to a laser amplifier as in (a). This panel then forms the basis for ``tiling'' the array. The actual number of optical elements in the module will depend on the final choice of laser amplifiers. (c) A portion of the array of hexagonal panels, as in (b), that make up the final array. The panels are as close packed as possible to minimize sidelobe power and be consistent with pointing requirements. Shown spacing allows 30 degree tilt.}
        \label{fig:laserdriver}
    \end{figure*}
    
    \begin{figure*}
        \centering
        \begin{tabular}{c c c}
             \includegraphics[width=0.45\textwidth]{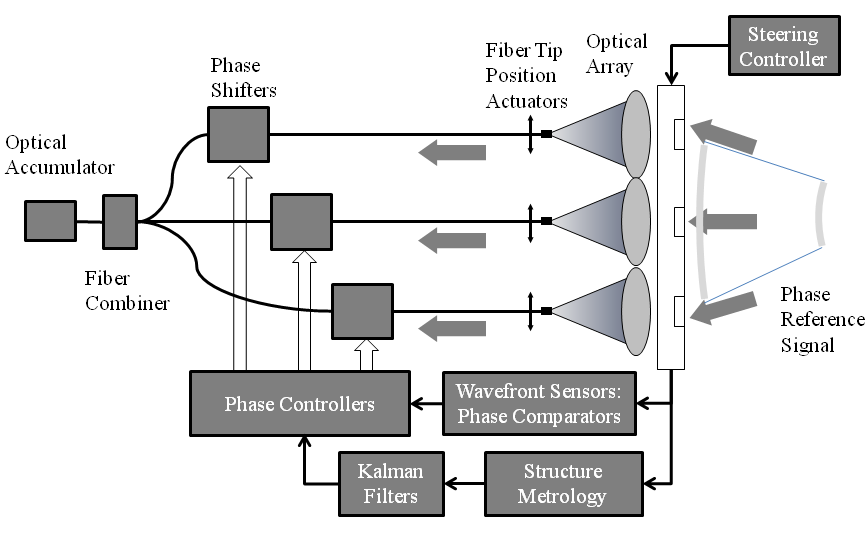} & \includegraphics[width=0.2\textwidth]{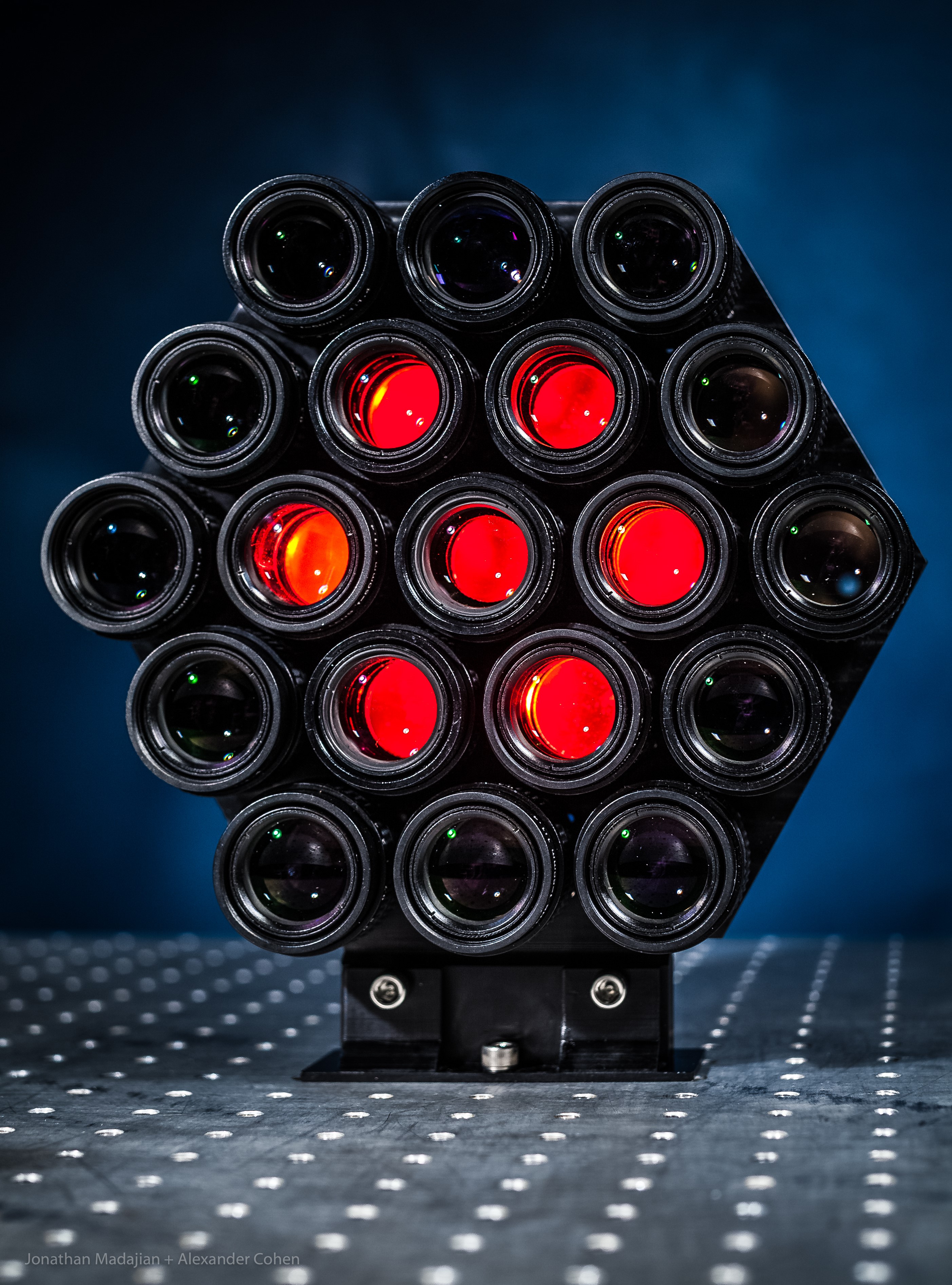} & \includegraphics[width=0.275\textwidth]{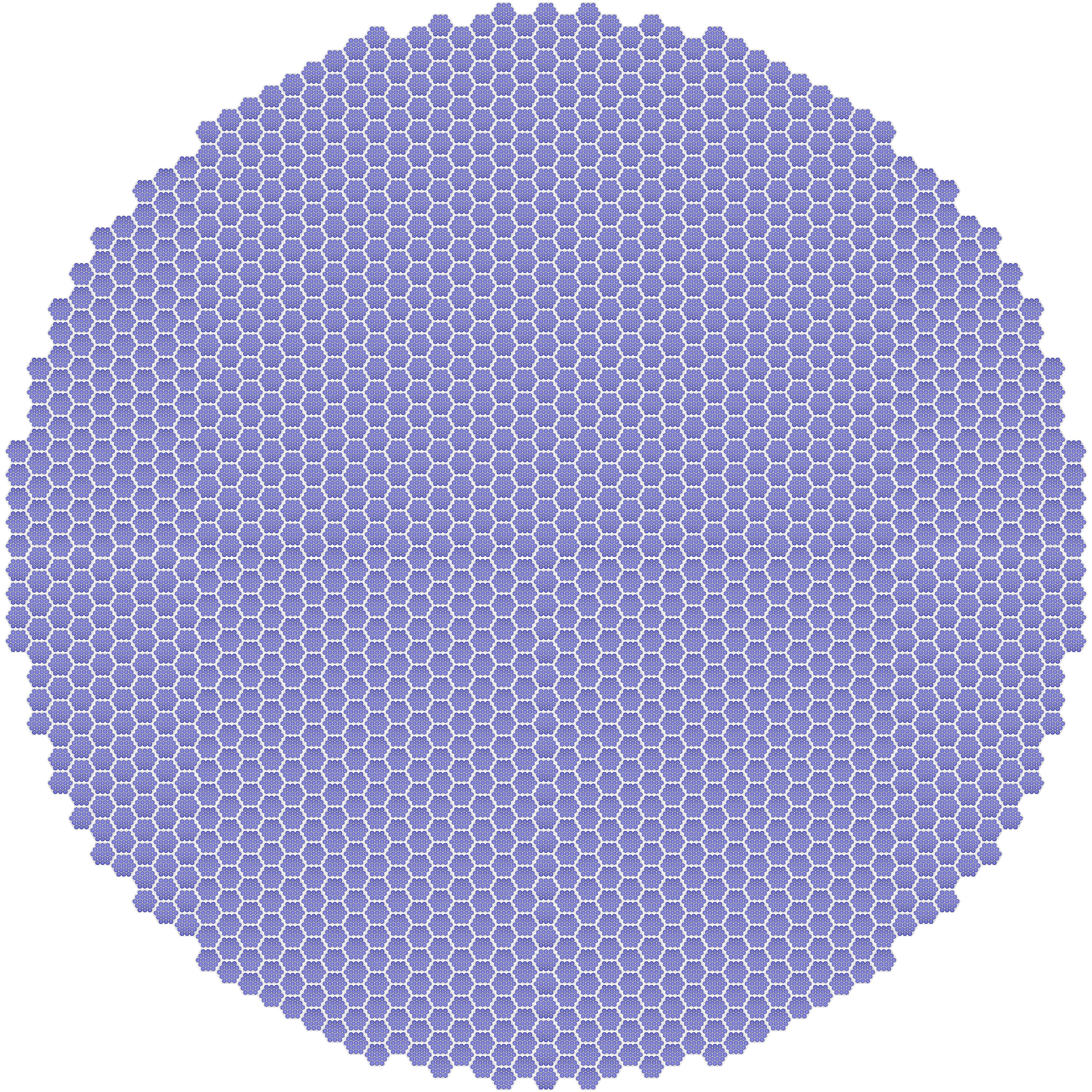}  \\
             (a) & (b) & (c) 
        \end{tabular}{}
        \caption{(a) The same laser array used for propulsion can be used in reverse as a phased array telescope allowing dual use of the system for large aperture laser communications reception (needed for interstellar missions) and offers a path forward to kilometer and large class telescopes for many other purposes. (b) 1/4 scale panel of phased array hexagonal module with single mode optical fibers attached. (c) Example of hexagonal close packed array -- diameter of 50 panels.}
        \label{fig:PAT}
    \end{figure*}
    
\section{Modularity and Scalability}

    The laser array (``photon engine'') is completely modular, scalable, and lends itself to mass production as all the elements are identical, as shown in Figure \ref{fig:laserdriver}. There are very large econ\-omies of scale in such a system in addition to the exponential growth. The system has no expendables, is completely solid state, and can run continuously for years on end. Industrial fiber lasers have MTBF in excess of 50,000 hours. The revolution in solid state lighting including upcoming laser lighting will only further increase the performance and lower  costs. The ``wall plug'' efficiency is excellent at 42\% as of this year. The same basic system can also be used as a phased array telescope for the receive side in the laser communications as well as for future kilometer-scale telescopes for specialized applications such as spectroscopy of exoplanet atmospheres and high redshift cosmology studies, as shown in Figure \ref{fig:PAT}.
    
\section{Exponential Growth in Photonics is Key}

\begin{figure*}
        \centering
        \begin{tabular}{c c c}
            \includegraphics[width=0.2\textwidth]{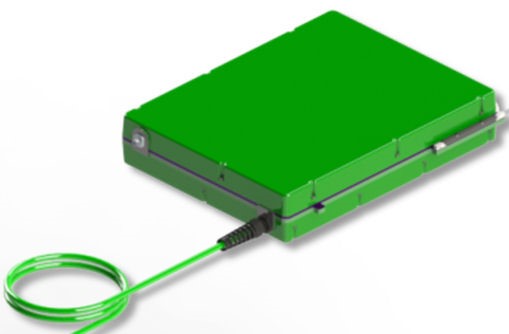} &
            \includegraphics[width=0.34\textwidth]{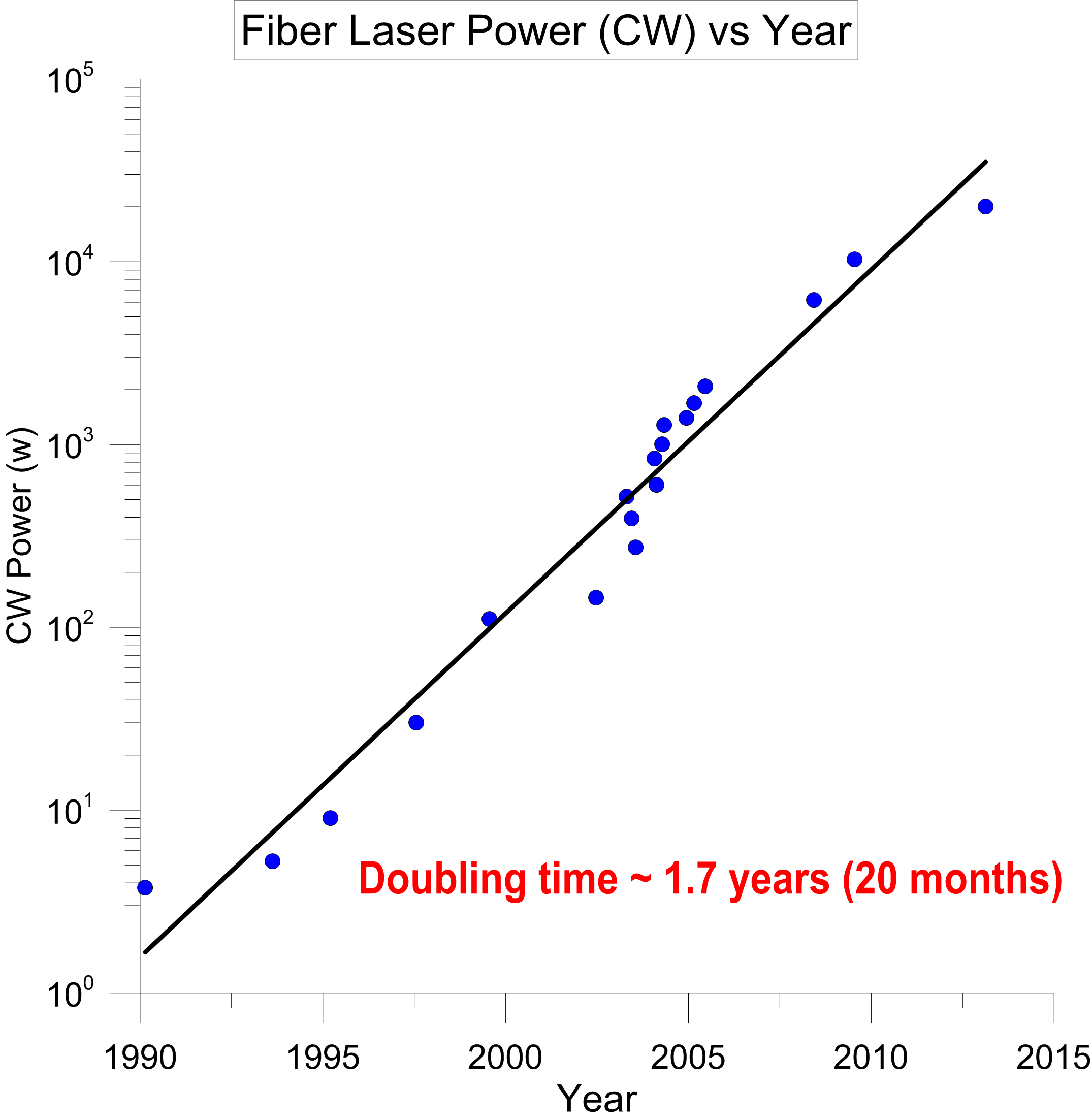} &
            \includegraphics[width=0.36\textwidth]{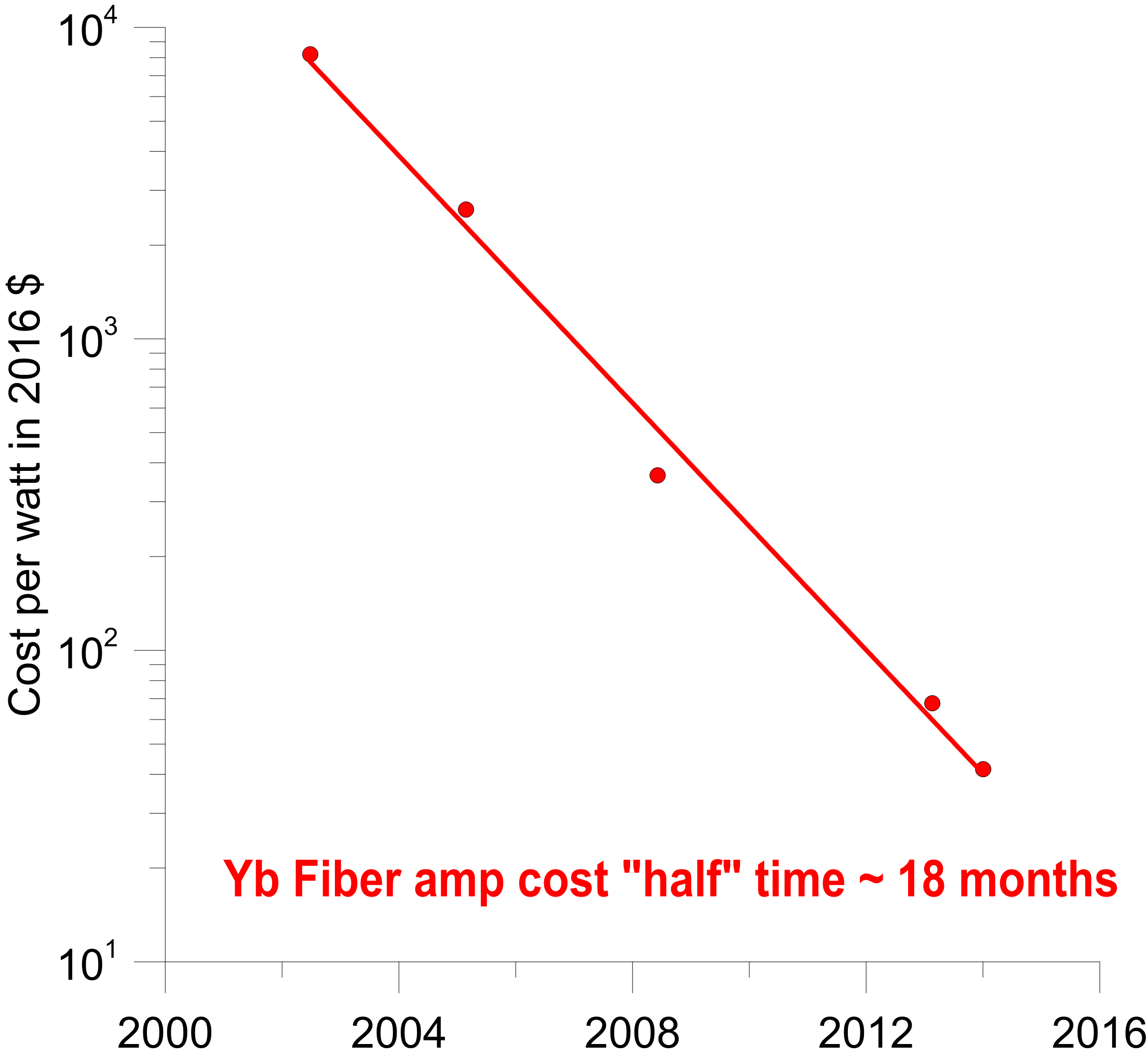} \\
            (a) & (b) & (c) \\
        \end{tabular}
        \caption{(a) Picture of current 1-3 kW class Yb laser amplifier which forms the baseline approach for our design. Fiber output is shown at lower left. Mass is approx 5 kg and size is approximately that of this page. This will evolve rapidly,            but is already sufficient to begin. Courtesy Nufern. (b) CW fiber laser power vs year over 25 years showing a ``Moore's Law'' like progression with a doubling time of about 20 months. (c) CW fiber lasers and Yb fiber laser amplifiers (baselined in this paper) cost/watt with an inflation index correction to bring it to 2016 dollars. Note the excellent fit to an exponential with a cost ``halving'' time of 18 months. }
        \label{fig:amplifiers}
    \end{figure*}

    Photonics, like electronics, is an exponential growth sector in both performance and in cost reductions with similar Moore's Law-like characteristics of doubling times in performance and halving times in cost of 18 months. This is radically different than chemical propulsion where performance has changed little since the dawn of the space age. This is shown clearly in Figure \ref{fig:amplifiers}. While there is wonderful innovation in chemical propulsion, ion engines, and related technologies, the innovation and drive in photonics is vastly beyond other propulsion technologies. This is largely due to the fundamentally low cost of the materials used and the ability to nanofabricate (wafer-scale) the relevant components. Much more development is coming in photonics and electronics in the coming decades, while performance in chemical propulsion peaked many decades ago. Of particular interest is the work in completely III-V on Si wafer-scale DE systems we are working on at UC Santa Barbara that could vastly lower the cost and increase the performance of the laser array.
\section{Cost Analysis}

    This program is enabled by the exponential growth in photonics as well as electronics which includes not only exponential growth in capability but also the exponential decrease in cost for a given metric (cost per unit power, for example). Given this exponential nature of the program it is critical to understand a detailed economic analysis of the system to determine which elements are exponential (such as photonics and electronics) and which are not (such as metals, concrete, and glass). We define cost modeling in terms of:
    \begin{itemize}
        \item[1)] Desired program outcome, such as mass and speed to be obtained.
        \item[2)] System parameters that can be changed, such as overall array power, array diameter, and wavelength.
    \end{itemize}
    The goal is then to minimize the cost for a given outcome, 1), by modifying the system parameters, 2). A significant complication is that any economic model dealing with exponential technological underpinning is extremely sensitive to time (wait longer and it is cheaper). On the other hand, when the technological underpinnings become sub-dominant in the overall cost, then other costs such as labor, land, and space launch may increase the cost with time. For example, until recently the cost of solar panels dominated the cost of a solar PV installation. That is no long true as secondary costs such as installation, wiring, inverters, and utility tie are now dominant. Then there is the ``cost of knowledge'' -- i.e. waiting delays knowledge.
    
    The other complication is that many system parameters are interconnected and there is the severe issue that we do not currently have the capacity to produce the required laser power levels we will need and hence industrial capacity will have to catch up, but we do not want to be the sole customer. Hence, finding technologies that are driven by other sectors or adopting technologies produced in mass quantity for other sectors may be required to get to the desired economic price point. For example, the current LED light market using GaN devices is currently less than \$0.1 per optical watt in large quantities and this is still dropping, though the cost of materials is starting to become a limit. IF we reach the same price point in our desired laser amplifiers in the next 20 years, as currently exists in LED lighting, then the overall system cost would NOT be dominated by the laser amplifier costs. Today the cost of (qty 1) laser amplifiers is $>$\$100 per optical watt or about 1000$\times$ larger than current LED lighting. However, a critical issue is that current LED costs are driven down in price by the volume used, whereas there is not such large volume in laser amplifiers of the type we want.
    
    Given all the above caveats, we do a simplified system analysis to get a general idea of cost optimization. In general, the cost functional $C_T$ is a multi-dimensional function of all the system parameters SP$_i$. Establishing a fully interconnected cost function is extremely difficult given the complex connectedness between the various parameters. In general, this system is not analytic in nature. Assuming we can make the cost function analytic and well behaved, we can in theory solve it to optimize the cost. If we can make it well behaved and differentiable then we would solve a series of coupled differential equations to search for a global minimum in the cost. With this assumption we have:
    \begin{equation}
        C_T(\textrm{DO}|\textrm{SP$_i$}),
    \end{equation}
    where
    \begin{align}
    \begin{split}
        \textrm{DO}\equiv&\hspace{0.5mm}\textrm{Desired Outcome -- (i.e. $v_0,m_0$)}\\
        \textrm{SP}_i\equiv&\hspace{0.75mm}\textrm{System Parameters -- (i.e. $P_0,d,\lambda,h,\rho,S_y$)}.
    \end{split}
    \end{align}
    Note that we immediately see the interconnections between system parameters. For example, the cost of a given power DOES depend on wavelength.
    
    To find a cost minimum we explore the multi-dimensional space and seek the minimum. We can do this by setting up a system of equations where the derivatives are all simultaneously set to zero. The various system parameters (SP) are interconnected via the Desired Outcomes (DO). This is WHY there is a minimum. It is the interconnection between the system parameters from the physics that relates SP to DO -- i.e. speed and mass are related to power, array size $d$, wavelength, etc. It is this deterministic relationship between SP and DO that gives a multi-dimensional form to the total cost. This results in a general solution ASSUMING the cost function is analytic. NOTE that the partial derivative DO NOT mean the other system parameters are held fixed during the differentiation since they may be interconnected from the physics of the desired outcome (DO).
    \begin{align}
    \begin{split}
        \frac{dC_T(\textrm{DO}|\textrm{SP$_i$})}{d\textrm{SP$_1$}}&=0\\
        \frac{dC_T(\textrm{DO}|\textrm{SP$_i$})}{d\textrm{SP$_2$}}&=0\\
        &\hspace{1.9mm}\vdots\\
        \frac{dC_T(\textrm{DO}|\textrm{SP$_i$})}{d\textrm{SP$_N$}}&=0.\\
    \end{split}
    \end{align}

\section{Independence of Costs of System Parameters Approximation}

    There are often elements of the system that are approximately independent of each other. For example, the laser amplifiers and optical elements (lenses or reflectors), structural elements, pointing systems, etc. can be approximately independent and individually costed. In this case the total cost of the system is the sum of the costs of these system elements. In this case the total cost is:
    \begin{equation}
        C_T=\sum_i C_i(\textrm{DO}|\textrm{SP$_i$}),
    \end{equation}
    where $C_i(\textrm{DO}|\textrm{SP$_i$})$ is the cost of the $i^{\textrm{th}}$ system element. The fraction of the total cost in the $i^{\textrm{th}}$ system element is:
    \begin{equation}
        f_i=\frac{C_i}{C_T},\hspace{1mm}\textrm{where}\hspace{1mm}\sum_i f_i=1.
    \end{equation}
    In this case, the set of coupled differential equations still applies, but it is now a set of coupled sums:
    \begin{align}
    \begin{split}
        \frac{dC_T(\textrm{DO}|\textrm{SP$_i$})}{d\textrm{SP$_1$}}&=\sum_i \frac{dC_i(\textrm{DO}|\textrm{SP$_i$})}{d\textrm{SP$_1$}}=0\\
        \frac{dC_T(\textrm{DO}|\textrm{SP$_i$})}{d\textrm{SP$_2$}}&=\sum_i \frac{dC_i(\textrm{DO}|\textrm{SP$_i$})}{d\textrm{SP$_2$}}=0\\
        &\hspace{1.8mm}\vdots\\
        \frac{dC_T(\textrm{DO}|\textrm{SP$_i$})}{d\textrm{SP$_N$}}&=\sum_i \frac{dC_i(\textrm{DO}|\textrm{SP$_i$})}{d\textrm{SP$_N$}}=0.\\
    \end{split}
    \end{align}
    
\section{Physics of SP$_i$ and DO}

    The physics relates the various system parameters and the desired outcomes. We showed previously the non-relativistic solution is as follows (reasonably accurate for $\beta_0<0.5$):
\subsection{Not materials strength limited case -- we choose sail thickness $h$}
\subsubsection{Not optimized case ($m_0\neq m_\textrm{ref}$):} 
    The speed at the point where the laser spot equals the sail size ($t=t_0$) is given by:
    \begin{align}
    \begin{split}
        v_0&=\bigg(\frac{P_0(1+\epsilon_r)dD}{c\lambda\alpha_d(\xi D^2h\rho+m_0)}\bigg)^{1/2}\\
        &\rightarrow \bigg(\frac{P_0(2\epsilon_r+(1-\epsilon_r)\alpha)dD}{c\lambda\alpha_d(\xi D^2h\rho+m_0)}\bigg)^{1/2}\\
        &=\bigg(\frac{P_0 \eta dD}{c\lambda\alpha_d(\xi D^2h\rho+m_0)}\bigg)^{1/2},
    \end{split}
    \end{align}
    where $\eta=(2\epsilon_r+(1-\epsilon_r)\alpha)$ as we allow both reflection and absorption. For a perfect reflector, we let $\eta=2$ and $\epsilon_r=1$. For a perfect absorber, we let $\eta=1$, $\epsilon_r=0$, and $\alpha=1$. We also have the relativistic $\beta$ factor at $t_0$:
    \begin{align}
    \begin{split}
        \beta_0&=\bigg(\frac{P_0(2\epsilon_r+(1-\epsilon_r)\alpha)dD}{c^3\lambda\alpha_d(\xi D^2h\rho+m_0)}\bigg)^{1/2}\\
        &=\bigg(\frac{P_0\eta dD}{c^3\lambda\alpha_d(\xi D^2h\rho+m_0)}\bigg)^{1/2}.
    \end{split}
    \end{align}
    The time at the distance where the laser spot equals the sail time is given by:
    \begin{align}
    \begin{split}
        t_0&=\frac{v_0}{a}=\bigg(\frac{cdD(\xi D^2h\rho+m_0)}{P_0(2\epsilon_r+(1-\epsilon_r)\alpha)\lambda\alpha_d}\bigg)^{1/2}\\
        &=\bigg(\frac{cdD(\xi D^2h\rho+m_0)}{P_0\eta\lambda\alpha_d}\bigg)^{1/2}
    \end{split}
    \end{align}
    and
    \begin{align}
    \begin{split}
        L_0=\frac{dD}{2\lambda\alpha_d}=\frac{d}{2\lambda\alpha_d}\bigg(\frac{m_\textrm{sail}}{\xi h\rho}\bigg)^{1/2}.
    \end{split}
    \end{align}
    The speed at infinity as limited by diffraction is given by:
    \begin{equation}
        v_\infty=\sqrt{2}v_0.
    \end{equation}
    
\subsubsection{Optimized case ($m_0=m_\textrm{ref}=\xi D^2 h\rho$):}
    \begin{align}
    \begin{split}
        v_0&=\bigg(\frac{P_0(2\epsilon_r+(1-\epsilon_r)\alpha)d}{2c\lambda\alpha_d}\bigg)^{1/2}(\xi h\rho m_0)^{-1/4}\\
        &=\bigg(\frac{P_0 \eta d}{2c\lambda\alpha_d}\bigg)^{1/2}(\xi h\rho m_0)^{-1/4}
    \end{split}
    \end{align}
    \begin{align}
    \begin{split}
        \beta_0&=\bigg(\frac{P_0(2\epsilon_r+(1-\epsilon_r)\alpha)d}{2c^3\lambda\alpha_d}\bigg)^{1/2}(\xi h\rho m_0)^{-1/4}\\
        &=\bigg(\frac{P_0 \eta d}{2c^3\lambda\alpha_d}\bigg)^{1/2}(\xi h\rho m_0)^{-1/4}.
    \end{split}
    \end{align}
    Note the only difference between the optimized case ($m_0=m_\textrm{ref}$) and the non-optimized case ($m_0\neq m_\textrm{ref}$) is that we replace
    \begin{equation}
        (\xi h\rho m_0)^{-1/4}
    \end{equation}
    with
    \begin{equation}
        \bigg(\frac{2D}{\xi D^2h\rho+m_0}\bigg)^{1/2}.
    \end{equation}
    For the optimized case we  have:
    \begin{align}
    \begin{split}
        t_0&=\bigg(\frac{2cdD^3\xi h\rho}{P_0(2\epsilon_r+(1-\epsilon_r)\alpha)\lambda\alpha_d}\bigg)^{1/2}\\
        &=\bigg(\frac{2cd}{P_0(2\epsilon_r+(1-\epsilon_r)\alpha)\lambda\alpha_d}\bigg)^{1/2}\bigg(\frac{m_0^3}{\xi h\rho}\bigg)^{1/4}\\
        &=\bigg(\frac{2cd}{P_0\eta\lambda\alpha_d}\bigg)^{1/2}\bigg(\frac{m_0^3}{\xi h\rho}\bigg)^{1/4}.
    \end{split}
    \end{align}
    
\subsection{Material strength-limited case -- material strength $S_y$\\ chooses sail thickness $h$:}

\subsubsection{Not optimized case ($m_0=m_\textrm{ref}$):}
    \begin{equation}
        D=2r=\frac{sP_0(2\epsilon_r+(1-\epsilon_r)\alpha)}{\pi hc S_y}=\frac{sP_0\eta}{\pi hc S_y},
    \end{equation}
    \begin{align}
    \begin{split}
        h&=\frac{sP_0(2\epsilon_r+(1-\epsilon_r)\alpha)}{2\pi rc S_y}\\
        &=\frac{sP_0(2\epsilon_r+(1-\epsilon_r)\alpha)}{\pi Dc S_y}\\
        &=\frac{sP_0\eta}{\pi Dc S_y}\sim\frac{1}{D},
    \end{split}
    \end{align}
    \begin{equation}
        v_0=\frac{P_0\eta}{c}\bigg(\frac{sd}{\pi hS_y\lambda\alpha_d(\xi D^2h\rho+m_0)}\bigg)^{1/2},
    \end{equation}
    \begin{equation}
        \rightarrow\beta_0=\frac{P_0\eta}{c^2}\bigg(\frac{sd}{\pi hS_y\lambda\alpha_d(\xi D^2h\rho+m_0)}\bigg)^{1/2}.
    \end{equation}
    
\subsubsection{Optimized case ($m_0=m_\textrm{ref}=\xi D^2 h\rho$):}

    With the optimized condition, we can compute $D$, $h$, and $v_0$ in terms of $m_0$:
    \begin{align}
    \begin{split}
        D&=\frac{m_\textrm{ref}}{\pi hD\rho}=\frac{m_0 cS_y}{\rho sP_0(2\epsilon_r+(1-\epsilon_r)\alpha)}\\
        &=\frac{m_0 cS_y}{\rho sP_0\eta}\propto m_0, P_0^{-1}, S_y, \rho^{-1}, \eta^{-1}
    \end{split}
    \end{align}
    \begin{align}
    \begin{split}
        h&=\frac{sP_0(2\epsilon_r+(1-\epsilon_r)\alpha)}{\pi DcS_y}\\
        &=\frac{\pi\rho}{m_0}\bigg(\frac{sP_0(2\epsilon_r+(1-\epsilon_r)\alpha)}{\pi cS_y}\bigg)^2\\
        &=\frac{\pi\rho}{m_0}\bigg(\frac{sP_0\eta}{\pi cS_y}\bigg)^2\propto m_0^{-1}, \rho, P_0^2, S_y^{-2}, \eta^2
    \end{split}
    \end{align}
    \begin{align}
    \begin{split}
        v_0&=\frac{P_0(2\epsilon_r+(1-\epsilon_r)\alpha)}{c}\bigg(\frac{sd}{2\pi hS_y\lambda\alpha_d m_0}\bigg)^{1/2}\\
        &=\frac{P_0\eta}{c}\bigg(\frac{sd}{2\pi hS_y\lambda\alpha_d m_0}\bigg)^{1/2}.
    \end{split}
    \end{align}
    In the case of no dependence on $\eta$, we have:
    \begin{equation}
        v_0=\bigg(\frac{dS_y}{2\rho\lambda\alpha_d s}\bigg)^{1/2}
    \end{equation}
    and
    \begin{equation}
        \beta_0=\bigg(\frac{dS_y}{2\rho c^2\lambda\alpha_d s}\bigg)^{1/2}\propto d^{1/2}, S_y^{1/2}, \rho^{-1/2}, \lambda^{-1/2}.
    \end{equation}
    
\subsection{A simple case}  
    
    A simple cost analysis case would be to assume the costs of each sub-element are independent and we assume the following:
    \begin{itemize}
        \item[1)] Choose wavelength from technological reasons
        \item[2)] Constant cost per watt
        \item[3)] Constant cost per unit area of optics including pointing elements
        \item[4)] Constant cost per unit energy used
        \item[5)] Constant cost per unit energy stored
        \item[6)] Constant cost of personnel
        \item[7)] For ground: constant cost per unit area of land
        \item[8)] For space: constant cost per unit mass launched and assembled
        \item[9)] Negligible cost per payload
    \end{itemize}{}
    For a simpler case we assume only 1) - 4) are important. We have the following definitions.
    
    Cost of total optical power (laser):
    \begin{equation}
        C_1=a_1 P_\textrm{optical}=\frac{a_1P_0}{\epsilon_b},\hspace{1mm}a_1=\$/\textrm{watt},
    \end{equation}
    where $\epsilon_b$ is the fraction of produced optical power in the main beam lobe:
    \begin{equation}
        \epsilon_b=\frac{P_0}{P_\textrm{optical}}.
    \end{equation}
    Cost of optical components (``glass cost''), dependent on array geometry ($\xi_\textrm{arr}=\pi/4$ for circular array, $\xi_\textrm{arr}=1$ for square):
    \begin{equation}
        C_2=a_2\xi d^2,\hspace{1mm}a_2=\$/\textrm{m$^2$}.
    \end{equation}
    Cost of energy used from the ``grid,'' including amortization of total number of missions:
    \begin{equation}
        C_3=a_3 P_0 t_0,\hspace{1mm}a_3=\$/\textrm{J}.
    \end{equation}
    Cost of energy stored, including amortization and efficiencies:
    \begin{equation}
        C_4=a_4 P_0 t_0,\hspace{1mm}a_4=\$/\textrm{J}.
    \end{equation}
    Note below in the physics that $t_0$ depends on $P_0$, $d$, $\lambda$, etc.
    \begin{align}
    \begin{split}
        \textrm{Fraction of cost in laser system:}\hspace{1mm}&f_1=\frac{C_1}{C_T},\\
        \textrm{Fraction of cost in optics system:}\hspace{1mm}&f_2=\frac{C_2}{C_T},\\
        \textrm{Fraction of cost in energy costs:}\hspace{1mm}&f_3=\frac{C_3}{C_T},\\
        \textrm{Fraction of cost in energy storage:}\hspace{1mm}&f_4=\frac{C_4}{C_T}.\\
    \end{split}
    \end{align}
    In this case:
    \begin{align}
    \begin{split}
        C_T&=C_1+C_2+C_3+C_4\\
        &=\frac{a_1P_0}{\epsilon_b}+a_2\xi_\textrm{arr}d^2+a_3P_0t_0+a_4P_0t_0.
    \end{split}
    \end{align}
    If we assume the cost of energy (dependent on the total number of missions over lifetime) and energy storage is small compared to the system cost, this simplifies to:
    \begin{align}
    \begin{split}
        C_T&=C_1+C_2=\frac{a_1P_0}{\epsilon_b}+a_2\xi_\textrm{arr}d^2=a_1P_\textrm{optical}+a_2\xi_\textrm{arr}d^2.
    \end{split}
    \end{align}
    In this case, we have only two system parameters that can be varied ($P$ and $d$) and these are NOT independent as they are related by the physics. This allows us to get a closed form analytic solution for the cost minimum. We would solve the following two equations:
    \begin{align}
    \begin{split}
        \diff{C_T}{P_0}&=\diff{C_1}{P_0}+\diff{C_2}{P_0}=0\\
        \diff{C_t}{d}&=\diff{C_1}{d}+\diff{C_2}{d}=0.
    \end{split}
    \end{align}
    There are two equations and two unknowns here ($P_0$ and $d$). $P_0$ and $d$ are related by the physics of the desired outcome (DO), as above. For the optimized case ($m_\textrm{sail}=m_0$) where we specify the sail thickness, we have:
    \begin{align}
    \begin{split}
        v_0&=\bigg(\frac{P_0(2\epsilon_r+(1-\epsilon_r)\alpha)}{2c\lambda\alpha_d}\bigg)^{1/2}(\xi h\rho m_0)^{-1/4}\\
        &=\bigg(\frac{P_0\eta d}{2c\lambda\alpha_d}\bigg)^{1/2}(\xi h\rho m_0)^{-1/4}.
    \end{split}
    \end{align}
    \begin{align}
    \begin{split}
        \therefore P_0&=v_0^2\bigg(\frac{2c\lambda\alpha_d}{(2\epsilon_r+(1-\epsilon_r)\alpha)d}\bigg)(\xi h\rho m_0)^{1/2}\\
        &=v_0^2\bigg(\frac{2c\lambda\alpha_d}{\eta d}\bigg)(\xi h\rho m_0)^{1/2}\\
        &=\beta_0^2\bigg(\frac{2c^3\lambda\alpha_d}{\eta d}\bigg)(\xi h\rho m_0)^{1/2}.
    \end{split}
    \end{align}
    The aperture flux (see figure below) is given by:
    \begin{align}
    \begin{split}
        F_\textrm{ap}(\textrm{W/m$^2$})=\frac{P_0}{\xi_\textrm{arr}d^2}&=\frac{2c^3\lambda\alpha_d (\xi h\rho m_0)^{1/2}}{(2\epsilon_r+(1-\epsilon_r)\alpha)\xi_\textrm{arr}d^3}\\
        &=\beta_0^2\frac{2c^3\lambda\alpha_d}{\eta\xi_\textrm{arr}d^3}(\xi h\rho m_0)^{1/2}.
    \end{split}
    \end{align}
    Therefore, the simplified cost of the system (ignoring energy and energy storage) is given by:
    \begin{align}
    \begin{split}
        C_T&=C_1+C_2=\frac{a_1P_0}{\epsilon_b}+a_2\xi_\textrm{arr}d^2\\
        &=\frac{a_1 v_0^2}{\epsilon_b}\frac{2c\lambda\alpha_d(\xi h\rho m_0)^{1/2}}{(2\epsilon_r+(1-\epsilon_r)\alpha)d}+a_2\xi_\textrm{arr}d^2\\
        &=\frac{a_1 v_0^2}{\epsilon_b}\frac{2c\lambda\alpha_d}{\eta d}(\xi h\rho m_0)^{1/2}+a_2\xi_\textrm{arr}d^2\\
        &=\frac{a_1'v_0^2}{d}+a_2'd^2=\frac{b_1'\beta_0^2}{d}+a_2'd^2\propto \frac{1}{d}+d^2,
    \end{split}
    \end{align}
    where
    \begin{align}
    \begin{split}
        a_1'&=\frac{a_1}{\epsilon_b}\frac{2c\lambda\alpha_d}{\eta}(\xi h\rho m_0)^{1/2}=0.600\frac{\lambda\alpha_d}{\eta}(\xi h\rho m_0)^{1/2},
    \end{split}
    \end{align}
    \begin{equation}
        a_2'=a_2\xi_\textrm{arr},
    \end{equation}
    and
    \begin{align}
    \begin{split}
        b_1'&=\frac{a_1}{\epsilon_b}\frac{2c^3\lambda\alpha_d}{\eta}(\xi h\rho m_0)^{1/2}\\
        &=5.40\times10^{16}\frac{a_1}{\epsilon_b}\frac{\lambda\alpha_d}{\eta}(\xi h\rho m_0)^{1/2},
    \end{split}
    \end{align}
    all for $[\lambda]=[h]=\mu$m, $[\rho]=$g/cc, and $[m_0]=$g.
    Therefore, the fraction cost in the laser system is given by:
    \begin{align}
    \begin{split}
        f_1=\frac{C_1}{C_T}=\frac{a_1'v_0^2/d}{a_1'v_0^2/d+a_2'd^2}&=\bigg(1+\frac{a_2'd^3}{a_1'v_0^2}\bigg)^{-1}\\
        &=\bigg(1+\frac{a_2'c^2\beta_0^2}{a_1'd^3}\bigg)^{-1}.
    \end{split}
    \end{align}
    As $d\rightarrow0$, then $f_1\rightarrow1$ and the laser system dominates the cost. The fraction of the total cost in the optics system is given by:
    \begin{align}
    \begin{split}
        f_2=\frac{C_2}{C_T}=\frac{a_2'd^2}{a_1'v_0^2/d+a_2'd^2}&=\bigg(1+\frac{a_1'v_0^2}{a_2'd^3}\bigg)^{-1}\\
        &=\bigg(1+\frac{a_1'c^2\beta_0^2}{a_2'd^3}\bigg)^{-1}.
    \end{split}
    \end{align}
    As $d\rightarrow\infty$, then $f_2\rightarrow1$ and the optics system dominates the cost.
    
    For this case of all costs in the laser (photonics) and optics (including mounting), $f_1+f_2=1$. We generalize to include energy and energy storage below.
    
    The $1/d+d^2$ terms are why there is a minimum cost point. For a desired speed, as the array size $d$ decreases, the power required goes up ($P_0 d$ is constant) and hence the cost of the laser amplifiers dominates ($1/d$ cost term), while at large array size the cost of optics dominates ($d^2$ term). There is an optimum as we show below. We have now reduced the case to depending on array size $d$.
    
    \begin{figure}[]
        \centering
        \begin{tabular}{c}
             \includegraphics[width=0.45\textwidth]{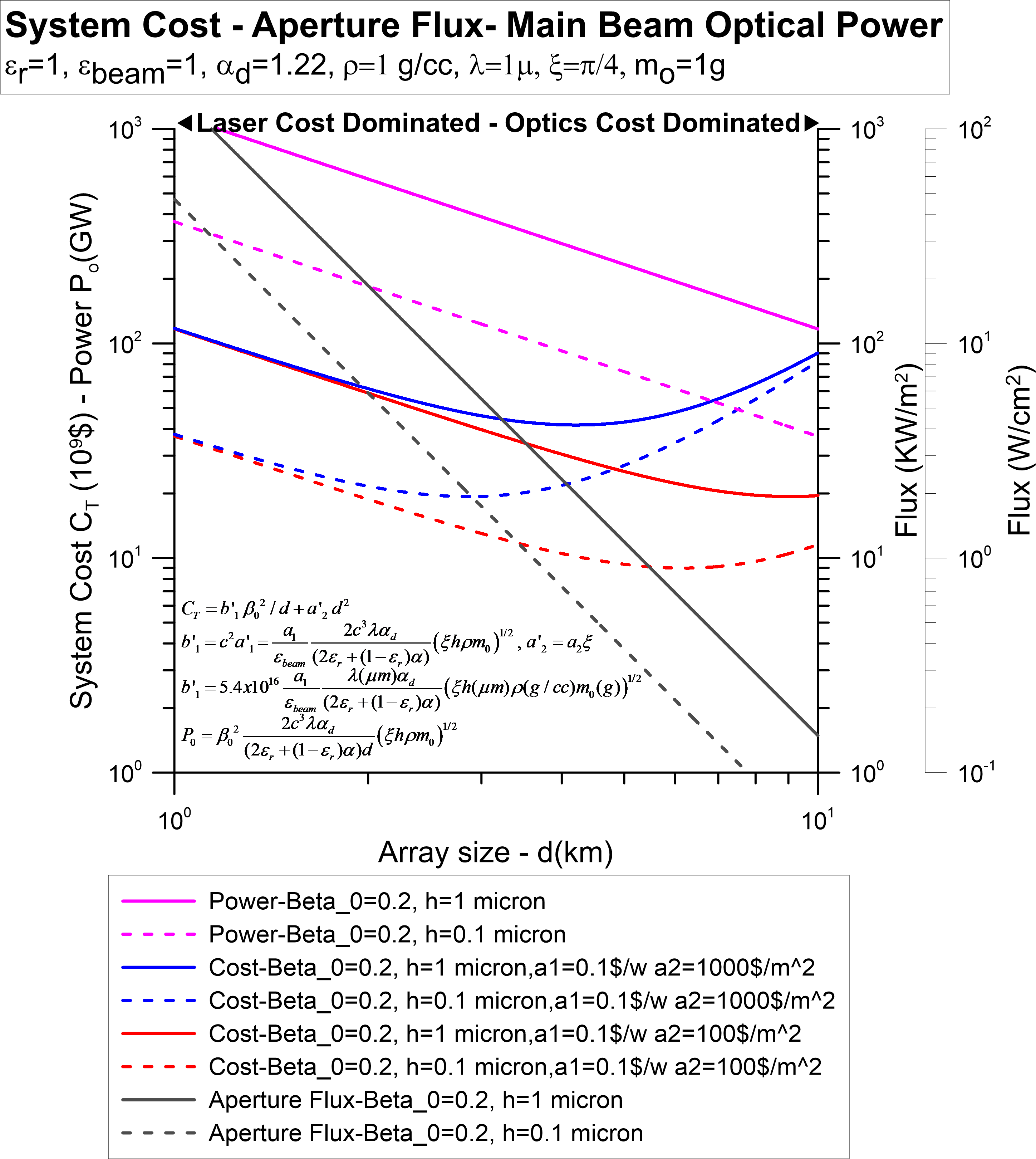}\\
             \includegraphics[width=0.45\textwidth]{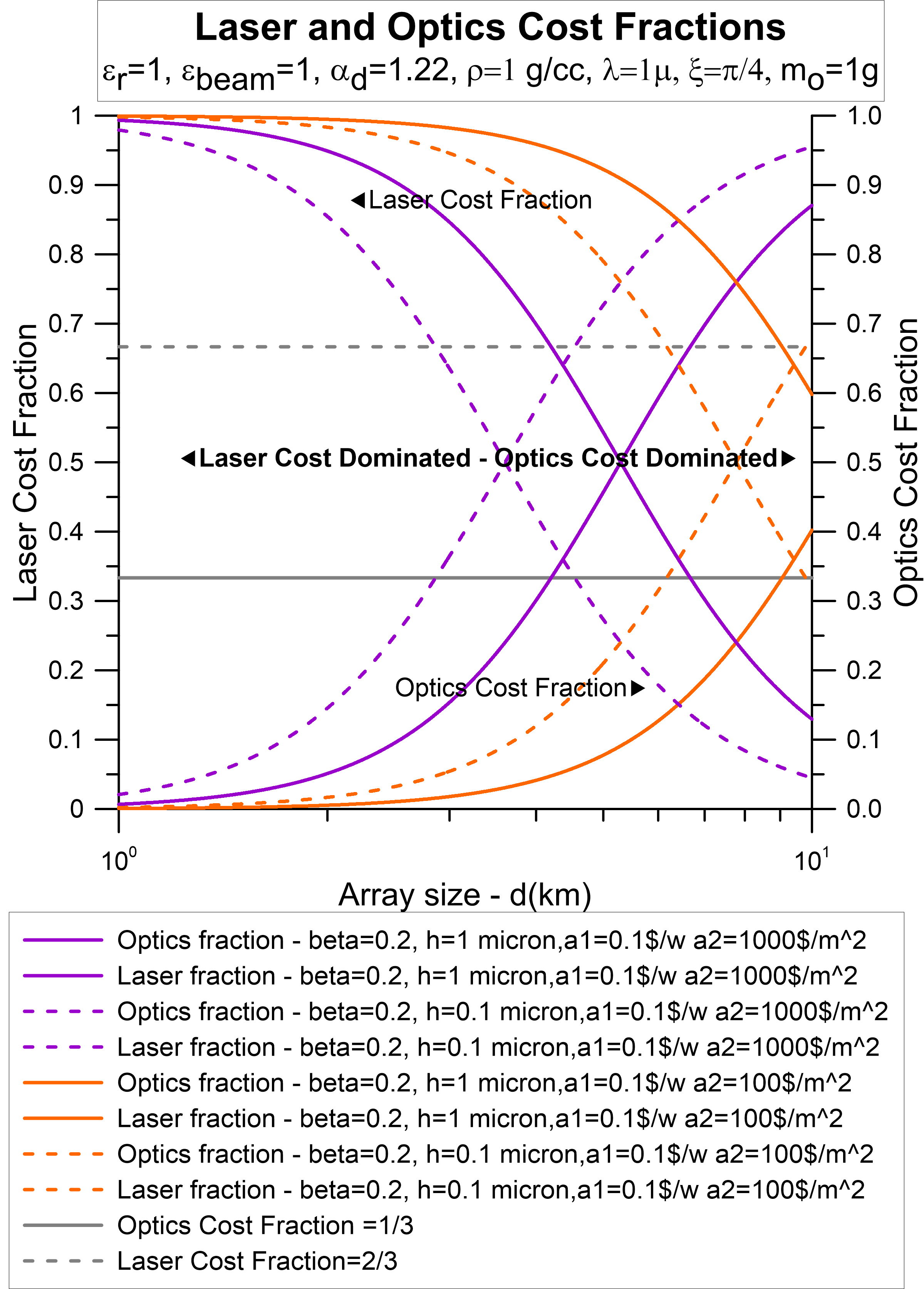}\\
        \end{tabular}{}
        \caption{(top) System cost $C_T$ as a function of array size $d$ in km for 1 and 0.1 micron sail thicknesses. Also shown is the main beam optical power in units of kW/m$^2$. (bottom)}
        \label{fig:systemcost}
    \end{figure}{}
    
    Hence, we get the minimum cost as follows:
    \begin{align}
    \begin{split}
        \diff{C_T}{d}&=0=-\frac{a_1 v_0^2}{\epsilon_b}\frac{2c\lambda\alpha_d}{\eta d^2}(\xi h\rho m_0)^{1/2}+2a_2\xi_\textrm{arr}\\
        &=a_2\bigg[\frac{-\beta_0a_1}{\epsilon_ba_2}\frac{2c^3\lambda\alpha_d}{\eta d^2}(\xi h\rho m_0)^{1/2}+2\xi_\textrm{arr}d\bigg].
    \end{split}
    \end{align}
    The minimum cost condition is then:
    \begin{equation}
        \frac{a_1v_0^2}{\epsilon_b}\frac{2c\lambda\alpha_d}{\eta d^2}(\xi h\rho m_0)^{1/2}=2a_2\xi_\textrm{arr}d.
    \end{equation}
    Note that the second derivative of the cost is always positive and thus the cost extreme is always a minimum:
    \begin{align}
    \begin{split}
        \diff[2]{C_T}{d}&=a_2\bigg[\frac{2\beta_0^2a_1}{\epsilon_b a_2}\frac{2c^3\lambda\alpha_d}{\eta d^3}(\xi h\rho m_0)^{1/2}+2\xi_\textrm{arr}\bigg]>0.
    \end{split}
    \end{align}
    The array size at the cost minimum is:
    \begin{align}
    \begin{split}
        d&=\bigg[\frac{a_1}{\epsilon_b}\frac{v_0^2}{2a_2\xi_\textrm{arr}}\frac{2c^3\lambda\alpha_d}{\eta d^3}(\xi h\rho m_0)^{1/2}\bigg]^{1/3}\\
        &=v_0^{2/3}\bigg(\frac{a_1}{\epsilon_b a_2}\bigg)^{1/3}\bigg[\frac{c\lambda\alpha_d}{\xi_\textrm{arr}\eta}(\xi h\rho m_0)^{1/2}\bigg]^{1/3}\\
        &=c\beta_0^{2/3}\bigg(\frac{a_1}{\epsilon_b a_2}\bigg)^{1/3}\bigg[\frac{\lambda\alpha_d}{\xi_\textrm{arr}\eta}(\xi h\rho m_0)^{1/2}\bigg]^{1/3}.
    \end{split}
    \end{align}
    For $\alpha_d=1.22$ and $\xi=\xi_\textrm{arr}=\pi/4$ (circle), we have:
    \begin{equation}
        d(m)\sim3.34\times10^5\beta_0^{2/3}\bigg(\frac{a_1}{\eta\epsilon_b a_2}\bigg)^{1/3}\Big(\lambda\sqrt{h\rho m_0}\Big)^{1/3}.
    \end{equation}
    Letting $\epsilon_r=1$ ($\eta=2$), we have:
    \begin{equation}
        d(m)\sim2.65\times10^5\beta_0^{2/3}\bigg(\frac{a_1}{\epsilon_b a_2}\bigg)^{1/3}\Big(\lambda\sqrt{h\rho m_0}\Big)^{1/3},
    \end{equation}
    for $[\lambda]=[h]=\mu$m, $[\rho]=$g/cc, and $[m_0]=$g.
    The results of the above calculations are plotted in Figure \ref{fig:systemcost} for 1 and 0.1 micron sail thicknesses. 
    
\subsection{Cost Minimum Depth}
     
\begin{figure}[]
        \centering
        \includegraphics[width=0.45\textwidth]{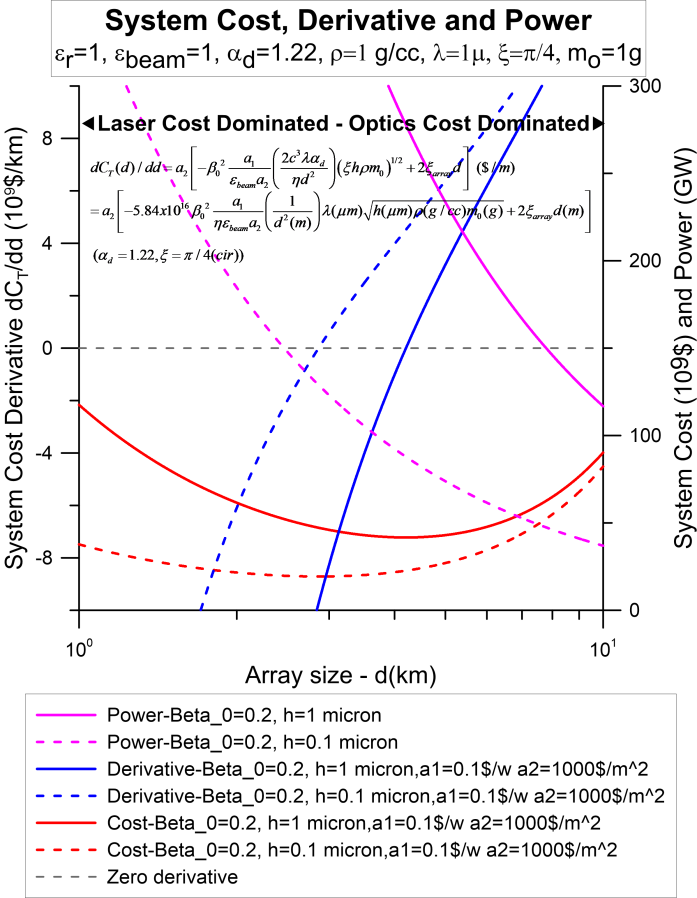}
        \caption{Array cost (red), cost derivative (blue), and power (magenta) as a function of array size for sail thickness $h=0.1$ and 1 micron.}
        \label{fig:costderivativepower}
    \end{figure}

    For many of the systems the cost minimum can be quite shallow as will be seen below in the system cost plots (Figure \ref{fig:costderivativepower}). We can explore this noting the cost sensitivity to changing the array size. The array main beam power at the cost minimum in the optimized case ($m_\textrm{ref}=m_0$) is given by:
    
    \begin{widetext}
    
    \begin{align}
    \begin{split}
        P_0(\textrm{W})&=v_0^2\bigg(\frac{2c\lambda\alpha_d}{\eta d}\bigg)(\xi h\rho m_0)^{1/2}=\beta_0^2\bigg(\frac{2c^3\lambda\alpha_d}{\eta d}\bigg)(\xi h\rho m_0)^{1/2}=\beta_0^2\bigg(\frac{2c^3\lambda\alpha_d}{\eta c\beta_0^{2/3}}\bigg)(\xi h\rho m_0)^{1/2}\bigg[\frac{a_1}{\epsilon_b a_2}\frac{\lambda\alpha_d}{\epsilon_\textrm{arr}a_2}(\xi h\rho m_0)^{1/2}\bigg]^{-1/3}\\
        &=\beta_0^{4/3}2c^2\bigg(\frac{a_1}{\epsilon_b a_2}\bigg)^{-1/3}\bigg(\frac{\alpha_d}{\eta}\bigg)^{2/3}\lambda^{2/3}\xi_\textrm{arr}^{1/3}(\xi h\rho m_0)^{1/3}=1.80\times10^{11}\beta_0^{4/3}\bigg(\frac{a_1}{\epsilon_b a_2}\bigg)^{-1/3}\bigg(\frac{\alpha_d}{\eta}\bigg)^{2/3}\lambda^{2/3}\xi_\textrm{arr}^{1/3}(\xi h\rho m_0)^{1/3}.
    \end{split}
    \end{align}
    For $\epsilon_r=1$ ($\eta=2$), $\alpha_d=1.22$, and $\xi=\xi_\textrm{arr}=\pi/4$ (circular array), we have:
    \begin{equation}
        P_0(\textrm{W})=1.10\times10^{11}\beta_0^{4/3}\bigg(\frac{a_1}{\epsilon_b a_2}\bigg)^{\hspace{-1mm}-1/3}\hspace{-2mm}\lambda^{2/3}(h\rho m_0)^{1/3},
    \end{equation}
    \begin{align}
    \begin{split}
        d(m)&=c\beta_0^{2/3}\bigg(\frac{a_1}{\epsilon_b a_2}\bigg)^{1/3}\bigg[\frac{\lambda\alpha_d}{\epsilon_\textrm{arr}\eta}(\xi h\rho m_0)^{1/2}\bigg]^{1/3}=3\times10^5\beta_0^{2/3}\bigg(\frac{a_1}{\epsilon_b a_2}\bigg)^{1/3}\bigg(\frac{\alpha_d}{\eta}\bigg)^{2/3}\lambda^{2/3}\xi_\textrm{arr}^{1/3}(\xi h\rho m_0)^{1/6}.
    \end{split}
    \end{align}
    Similarly, for $\epsilon_r=1$ ($\eta=2$), $\alpha_d=1.22$, and $\xi=\xi_\textrm{arr}=\pi/4$ (circular array), we have:
    \begin{equation}
        d(m)=2.65\times10^5\beta_0^{2/3}\bigg(\frac{a_1}{\epsilon_b a_2}\bigg)^{1/3}\lambda^{1/3}(h\rho m_0)^{1/6}.
    \end{equation}
    The cost minimum for no energy storage is $C_T=C_1(\textrm{laser})+C_2(\textrm{optics})$, where
    \begin{align}
    \begin{split}
        C_1&=\frac{a_1 P_0}{\epsilon_b}=\frac{a_1}{\epsilon_b}\beta_0^{4/3}2c^2\bigg(\frac{a_1}{\epsilon_b a_2}\bigg)^{-1/3}\bigg(\frac{\alpha_d}{\eta}\bigg)^{2/3}\lambda^{2/3}\xi_\textrm{arr}^{1/3}(\xi h\rho m_0)^{1/3}=a_1^{2/3}a_2^{1/3}\epsilon_b^{-2/3}\beta_0^{4/3}2c^2\bigg(\frac{\alpha_d}{\eta}\bigg)^{2/3}\lambda^{2/3}\xi_\textrm{arr}^{1/3}(\xi h\rho m_0)^{1/3}\\
        &=1.80\times10^{11}\beta_0^{4/3}\frac{a_1}{\epsilon_b}\bigg(\frac{a_1}{\epsilon_b a_2}\bigg)^{-1/3}\bigg(\frac{\alpha_d}{\eta}\bigg)^{2/3}\lambda^{2/3}\xi_\textrm{arr}^{1/3}(\xi h\rho m_0)^{1/3}=1.80\times10^{11}\beta_0^{4/3}a_1^{2/3}a_2^{1/3}\epsilon_b^{-4/3}\bigg(\frac{\alpha_d}{\eta}\bigg)^{2/3}\lambda^{2/3}\xi_\textrm{arr}^{1/3}(\xi h\rho m_0)^{1/3},
    \end{split}
    \end{align}
    and
    \begin{align}
    \begin{split}
        C_2=a_2\xi_\textrm{arr}d^2=a_2\xi_\textrm{arr}c^2\beta_0^{4/3}\bigg(\frac{a_1}{\epsilon_b a_2}\bigg)^{2/3}\bigg[\frac{\lambda\alpha_d}{\xi_\textrm{arr}\eta}(\xi h\rho m_0)^{1/2}\bigg]^{2/3}&=a_1^{2/3}a_2^{1/3}\epsilon_b^{-2/3}c^2\beta_0^{4/3}\bigg(\frac{\alpha_d}{\eta}\bigg)^{2/3}\lambda^{2/3}\xi_\textrm{arr}^{1/3}(\xi h\rho m_0)^{1/3}\\
        &=9.00\times10^{10}\beta_0^{4/3}a_1^{2/3}a_2^{1/3}\epsilon_b^{-4/3}\bigg(\frac{\alpha_d}{\eta}\bigg)^{2/3}\lambda^{2/3}\xi_\textrm{arr}^{1/3}(\xi h\rho m_0)^{1/3}.
    \end{split}
    \end{align}
    For $\epsilon_r=1$ ($\eta=2$), $\alpha_d=1.22$, and $\xi=\xi_\textrm{arr}=\pi/4$ (circular array), we have:
    \begin{align}
    \begin{split}{}
        C_1&=1.10\times10^{11}\beta_0^{4/3}a_1^{2/3}a_2^{1/3}\epsilon_b^{-4/3}\lambda^{2/3}(h\rho m_0)^{1/3}\\
        C_2&=5.50\times10^{10}\beta_0^{4/3}a_1^{2/3}a_2^{1/3}\epsilon_b^{-4/3}\lambda^{2/3}(h\rho m_0)^{1/3}.
    \end{split}
    \end{align}
    Therefore, we see that:
    \begin{align}
    \begin{split}
        C_1=2C_2\rightarrow C_T&=C_1+C_2=3C_2=1.5C_1=2.70\times10^{11}\beta_0^{4/3}a_1^{2/3}a_2^{1/3}\epsilon_b^{-4/3}\bigg(\frac{\alpha_d}{\eta}\bigg)^{2/3}\lambda^{2/3}\xi_\textrm{arr}^{1/3}(\xi h\rho m_0)^{1/3},
    \end{split}
    \end{align}
    or for $\epsilon_r=1$ ($\eta=2$), $\alpha_d=1.22$, and $\xi=\xi_\textrm{arr}=\pi/4$ (circular array), we have:
    \begin{equation}
        C_T=1.65\times10^{11}\beta_0^{4/3}a_1^{2/3}a_2^{1/3}\epsilon_b^{-4/3}\lambda^{2/3}(h\rho m_0)^{1/3}.
    \end{equation}
    \end{widetext}
    
    {}
    
    Note the system cost and pricing for the laser and optics at the minimum cost with no storage all scale as:
    \begin{equation}
        C_T\propto\beta_0^{4/3}a_1^{2/3}a_2^{1/3}.
    \end{equation}
    Note the power law scaling of 2/3 and 1/3 respectively for the laser ($a_1$) and optics ($a_2$) cost metrics. The power indices of 2/3 and 1/3 are the reason for the laser and optics costs being 2/3 and 1/3 of the total cost.
    
    The laser power ($P_0$) and array size ($d$) scale as:
    \begin{align}
    \begin{split}
        P_0&=\beta_0^{4/3}2c^2\bigg(\frac{a_1}{\epsilon_b a_2}\bigg)^{\hspace{-1.5mm}-1/3}\hspace{-1mm}\bigg(\frac{\alpha_d\lambda}{\eta}\bigg)^{\hspace{-1mm}2/3}\hspace{-1mm}(\xi_\textrm{arr}\xi h\rho m_0)^{1/3}\\
        &\propto\beta_0^{4/3}a_1^{-1/3}a_2^{1/3}
    \end{split}
    \end{align}
    \begin{align}
    \begin{split}
        d(m)&=c\beta_0^{2/3}\bigg(\frac{a_1}{\epsilon_b a_2}\bigg)^{1/3}\bigg[\frac{\lambda\alpha_d}{\xi_\textrm{arr}\eta}(\xi h\rho m_0)^{1/2}\bigg]^{1/3}\\
        &\propto\beta_0^{2/3}a_1^{1/3}a_2^{-1/3}.
    \end{split}
    \end{align}
    
    For $\epsilon_r=1$ ($\eta=2$), $\alpha_d=1.22$, and $\xi=\xi_\textrm{arr}=\pi/4$ (circular array), we can explicitly calculate the minimum total cost $C_T$ given fixed system parameters ($\textrm{SP}_i$) and desired outcomes (DO):
    \begin{itemize}
        \item[1)] $\beta_0=0.2$, $m_0=1$ g, $h=1$ $\mu$m, $\rho=1$ g/cc, $\lambda=1$ $\mu$m, $a_1/\epsilon_b=$ $\$1/W$, $a_2=$ $\$1000$/m$^2$ $\rightarrow$ $d=9.1$ km, $P_0=128$ GW and minimum total cost is $C_T\sim$ \$193B.
        \item[2)] $\beta_0=0.2$, $m_0=1$ g, $h=1$ $\mu$m, $\rho=1$ g/cc, $\lambda=1$ $\mu$m, $a_1/\epsilon_b=$ $\$0.1/W$, $a_2=$ $\$1000$/m$^2$ $\rightarrow$ $d=4.2$ km, $P_0=272$ GW and minimum total cost is $C_T\sim$ \$41B.
    \end{itemize}{}
    Once we find the array size $d$ we then can solve for the power $P_0$ as well as the total cost $C_T$. Notice that as we drive $a_1$ down (cost per watt goes down) the array size decreases, while if we drive $a_2$ down (cost per m$^2$ of optics) then $d$ increases. This tradeoff is very logical as we trade laser power costs for array size. This is the logical consequence of the speed depending on the $P_0$ and is seen in the physics above.
    
\subsection{Understanding the Cost Minimum}

    Some physical insight into the nature of the cost minimum is hidden in the mathematics, but it becomes clear if we realize the following:
    \begin{align}
    \begin{split}
        \diff{C_T}{d}&=\diff{C_1}{d}+\diff{C_2}{d}=0\\
        &=\diff{}{d}\bigg(\frac{a_1 P_0}{\epsilon_b}\bigg)+\diff{}{d}(a_2\xi_\textrm{arr}d^2)\\
        &=\frac{a_1}{\epsilon_b}\diff{P_0}{d}+2a_2\xi_\textrm{arr}d.
    \end{split}
    \end{align}
    Recalling that
    \begin{equation}
        P_0=v_0^2\bigg(\frac{2c\lambda\alpha_d}{(2\epsilon_r+(1-\epsilon_r)\alpha)d}\bigg)(\xi h\rho m_0)^{1/2},
    \end{equation}
    we have:
    \begin{equation}
        \diff{P_0}{d}=-\frac{P_0}{d}.
    \end{equation}
    This is due to the functional form of $P_0(d)$.
    Thus,
    \begin{align}
    \begin{split}
        \diff{C_T}{d}&=\frac{a_1}{\epsilon_b}\diff{P_0}{d}+2a_2\xi_\textrm{arr}d=0\\
        &=-\frac{a_1}{\epsilon_b}\frac{P_0}{d}+2a_2\xi_\textrm{arr}d,
    \end{split}
    \end{align}
    and therefore,
    \begin{equation}
        \frac{a_1}{\epsilon_b}P_0=a_1P_\textrm{optical}=2a_2\xi_\textrm{arr}d^2.
    \end{equation}
    This implies that the cost of the total optical power (laser amplifiers) is equal to $2\times$ the cost of the optics. The bottom line is that in this simple case of separable laser amplifier costs and optics costs as the only two items, the minimum cost is always when:
    \begin{equation}
        (\textrm{Cost of optical power})=2\times(\textrm{Cost of optics}).
    \end{equation}
    In terms of the total cost, we have:
    \begin{align}
    \begin{split}
        C_T&=1.5\times(\textrm{Cost of total optical power})\\
        C_T&=3\times(\textrm{Cost of optics}),
    \end{split}
    \end{align}
    where the total optical power $P_T=P_0/\epsilon_b$.
    
    This is all because of the functional form of the relationship between the optical power $P_0$ and the array size $d$ (this is from the physics of the problem) and the assumed functional form of the optics cost (proportional to area or $d^2$). These combine to give us a cost minimum (if these are the only two or dominant costs) so that the cost of the optical power is twice that of the optics.
    
\subsection{Materials Strength Limited vs. Manufacturing Limits}

    Notice that in the above analysis we can either specify the sail thickness $h$ or work from a materials strength limited reg\-ime by specifying $S_y$ which then gives us $h$. In general, this may be a manufacturing limit (i.e. we would like single layer graphene but cannot produce it in the sizes and reflectivity we want) or it may be a materials strength limit (we need thicker material to withstand the stresses). In the optimized ($m_\textrm{ref}=m_0$) material strength limited case, we get $D$ and $h$ as follows:
    \begin{align}
    \begin{split}
        D=\frac{m_\textrm{ref}}{\pi hD\rho}&=\frac{m_0 cS_y}{\rho sP_0(2\epsilon_r+(1-\epsilon_r)\alpha)}\propto m_0, P_0^{-1}, S_y, \rho^{-1}.
    \end{split}
    \end{align}
    \begin{align}
    \begin{split}
        h&=\frac{sP_0(2\epsilon_r+(1-\epsilon_r)\alpha)}{\pi DcS_y}\\
        &=\frac{\pi\rho}{m_0}\bigg[\frac{sP_0(2\epsilon_r+(1-\epsilon_r)\alpha)}{\pi cS_y}\bigg]^2.
    \end{split}
    \end{align}
    
\section{Fixed Cost Optimization of Speed}

    For the same assumptions above of system cost dominated by only the laser and the optics, we can also analyze a case of fixed cost and maximizing the speed for a given spacecraft mass and set of sail parameters.  We proceed as follows using the cost analysis above for the optimized case of $m_\textrm{ref}=m_0$:
    \begin{align}
    \begin{split}
        C_T&=C_1+C_2=\frac{a_1}{\epsilon_b}P_0+a_2\xi_\textrm{arr}d^2\\
        &=\frac{a_1}{\epsilon_b}v_0^2\bigg(\frac{2c\lambda\alpha_d}{\eta d}\bigg)(\xi h\rho m_0)^{1/2}+a_2\xi_\textrm{arr}d^2.
    \end{split}
    \end{align}
    \begin{align}
    \begin{split}
        \therefore v_0^2&=\bigg[\frac{C_T-a_2\xi_\textrm{arr}d^2}{a_1/\epsilon_b}\frac{\eta d}{2c\lambda\alpha_d(\xi h\rho m_0)^{1/2}}\bigg]\\
        &=\frac{\eta \big(C_T d-a_2\xi_\textrm{arr}d^3\big)}{2c\lambda\alpha_d(a_1/\epsilon_b)(\xi h\rho m_0)^{1/2}}.
    \end{split}
    \end{align}
    \begin{align}
    \begin{split}
        \therefore\beta_0^2=\frac{\eta\big(C_T d-a_2\xi_\textrm{arr}d^3\big)}{2c^3\lambda\alpha_d(a_1/\epsilon_b)(\xi h\rho m_0)^{1/2}}.
    \end{split}
    \end{align}
    Note the dependence of speed above upon the total cost $C_T$ and array size $d$. The term $(C_T d-a_2\xi d^3)$ in the speed is interpreted as follows:
    \begin{itemize}
        \item Small $d$ $\rightarrow$ all the cost is spent on the laser $\rightarrow$ higher power, smaller aperture $\rightarrow$ lower speed.
        \item As $d$ starts to increase, speed increases due to longer range of illumination.
        \item Large $d$ $\rightarrow$ costs are increasingly spent on optics $\rightarrow$ lower power (less money for laser).
        \item As $d$ grows large the amount of power available (less money) is much less $\rightarrow$ lower speed.
        \item Hence, there is a maximum in the speed vs. array size function.
    \end{itemize}{}
    Maximizing $v_0$ is equivalent to maximizing $v_0^2$ or $\beta_0^2$:
    \begin{align}
    \begin{split}
        \diff{v_0^2}{d}&=\diff{}{d}\big(C_T d-a_2\xi_\textrm{arr}d^3\big)\\
        0&=C_T-3a_2\xi_\textrm{arr}d^2\\
        a_2\xi_\textrm{arr}d^2&=C_T/3.
    \end{split}
    \end{align}
    Therefore, as we derived before for the case of minimum cost for fixed $v_0$, the maximum speed occurs when the optics cost is 1/3 of the total cost. The maximum speed for fixed cost $C_T$ is thus when:
    \begin{equation}
        d=\bigg[\frac{C_T}{3a_2\xi_\textrm{arr}}\bigg]^{1/2}
    \end{equation}
    \begin{align}
    \begin{split}{}
        P_0&=v_0^2\bigg(\frac{2c\lambda\alpha_d}{\eta d}\bigg)(\xi h\rho m_0)^{1/2}\\
        &=\beta_0^2\bigg(\frac{2c^3\lambda\alpha_d}{\eta d}\bigg)(\xi h\rho m_0)^{1/2},
    \end{split}
    \end{align}
    \begin{align}
    \begin{split}
        \beta_0^2&=\frac{\eta}{2c^3\lambda\alpha_d a_1(\xi h\rho m_0)^{1/2}}\big(C_T d-a_2\xi_\textrm{arr}d^3\big)\\
        &=\frac{\eta}{2c^3\lambda\alpha_d a_1(\xi h \rho m_0)^{1/2}}d\big(C_T-a_2\xi_\textrm{arr}d^2\big)\\
        &=\frac{\eta}{2c^3\lambda\alpha_d a_1(\xi h\rho m_0)^{1/2}}d\frac{2C_T}{3}.
    \end{split}
    \end{align}
    
    When the case is material strength limited and when optimized ($m_\textrm{ref}=m_0$), the reflector diameter and thickness are given by Equations 69 and 70.
    
\subsection{Relationship Between $a_1$ and $a_2$ for Fixed $C_T$ and $\beta_0$ for the Minimum System Cost}

\begin{figure}[h]
        \centering
        \includegraphics[width=0.45\textwidth]{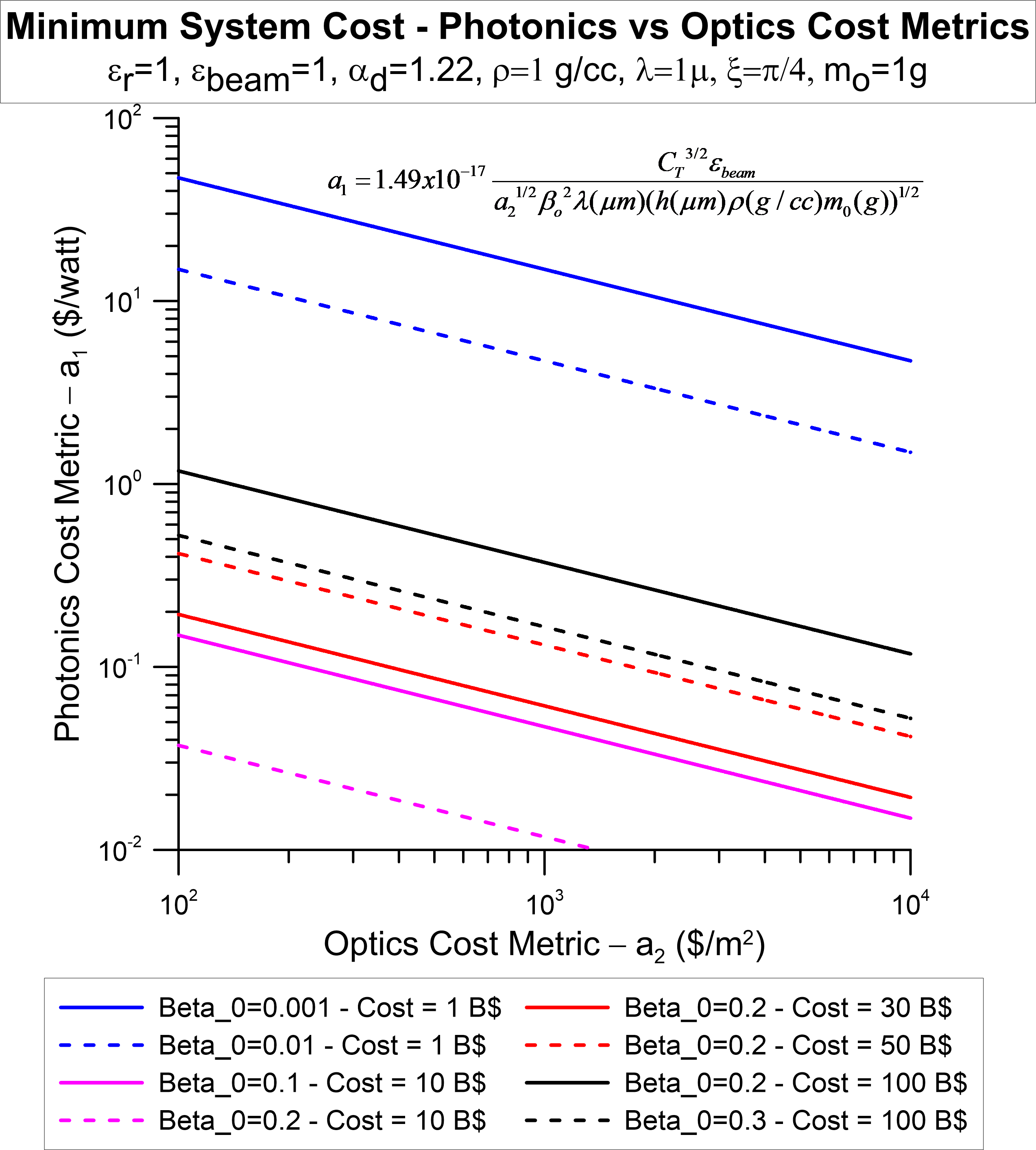}
        \caption{Relationship between the photonics and optics cost metrics, $\alpha_1$ and $\alpha_2$, for a variety of spacecraft $\beta$.}
        \label{fig:minimumcost}
    \end{figure}

    We can solve for the relationship between the cost metric for the laser and optics, $a_1$ and $a_2$, to give the minimum cost given a fixed minimum cost $C_T$ and $\beta_0$:
        \begin{widetext}
    \begin{align}
    \begin{split}
        C_T&=C_1+C_2=3C_2=1.5C_1=2.70\times10^{11}\beta_0^{4/3}a_1^{2/3}a_2^{1/3}\epsilon_b^{-4/3}\bigg(\frac{\alpha_d}{\eta}\bigg)^{2/3}\lambda^{2/3}\xi_\textrm{arr}^{1/3}(\xi h\rho m_0)^{1/3}.
    \end{split}
    \end{align}
    For $\epsilon_r=1$ ($\eta=2$), $\alpha_d=1.22$, and $\xi=\xi_\textrm{arr}=\pi/4$ (circular array), we have:
    \begin{equation}
        C_T=1.65\times10^{11}\beta_0^{4/3}a_1^{2/3}a_2^{1/3}\epsilon_b^{-4/3}\lambda^{2/3}(h\rho m_0)^{1/3},
    \end{equation}
    \begin{align}
    \begin{split}
        a_1^2 a_2=2.23\times10^{-34}\frac{C_T^3\epsilon_b^2}{\beta_0^4\lambda^2(h\rho m_0)}\rightarrow a_1=1.49\times10^{-17}\frac{C_T^{3/2}\epsilon_b}{a_2^{1/2}\beta_0^2\lambda(h\rho m_0)^{1/2}}.
    \end{split}
    \end{align}
    \end{widetext}
    In general, the laser (photonics) costs are governed by the exponential photonics technology and will drop significantly with time while the optics (``glass'') are not an exponential technology. For a given optics cost $a_2$ (per unit area) we can compute the required $a_1$, as shown in Figure \ref{fig:minimumcost}. We can see that for a fixed cost $C_T$ that the required $a_1\sim1/\beta_0^2$ and hence achieving an increasing $\beta_0$ requires a rapidly reduced $a_1$ to achieve a fixed price goal. 
    
    {}
    
\section{A Staged Development Approach}

\begin{figure}[]
        \centering
        \includegraphics[width=0.45\textwidth]{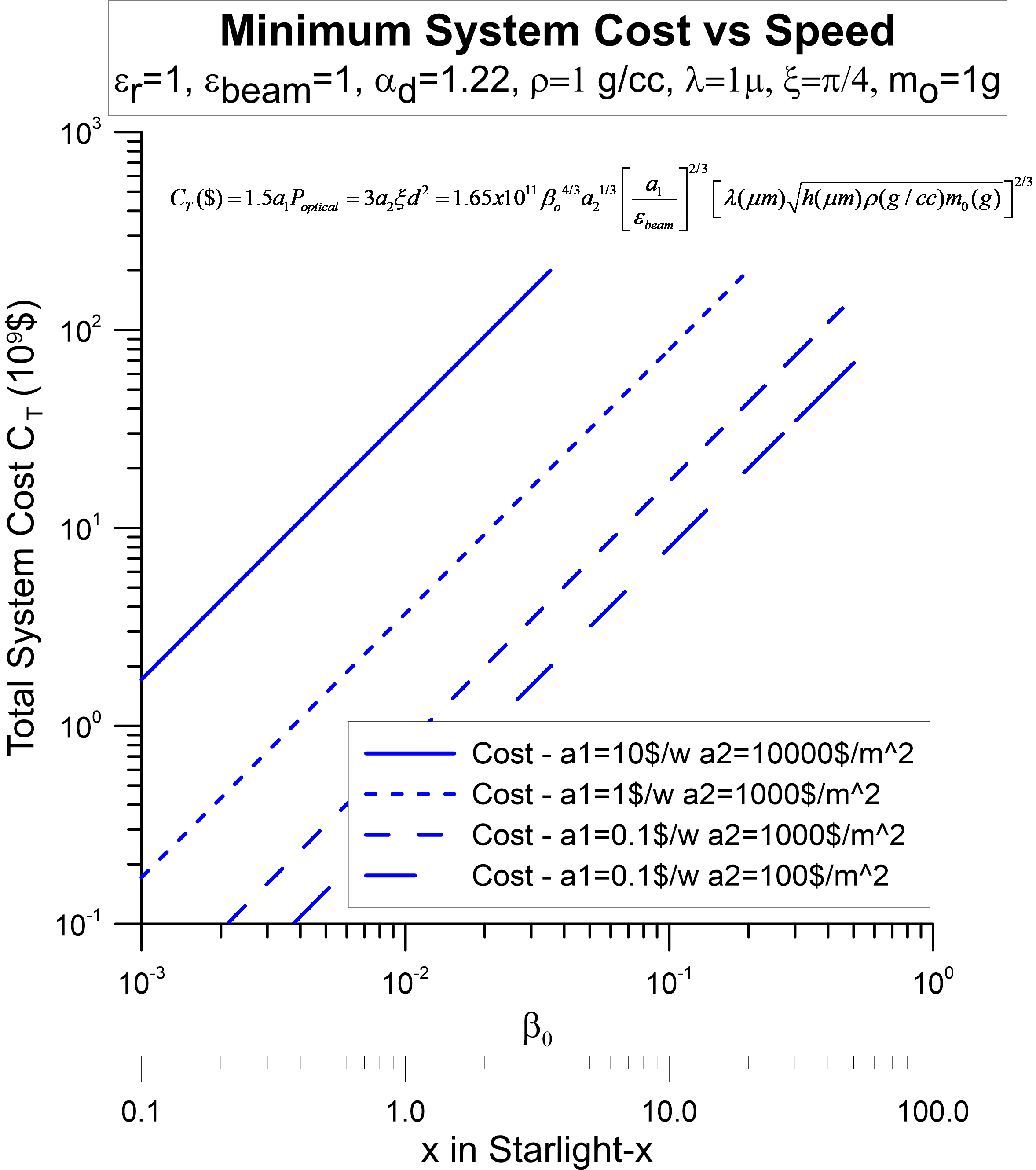}
        \caption{Minimum total system cost as a function of spacecraft $\beta$.}
        \label{fig:mincostvsspeed}
    \end{figure}

    Along the roadmap to interstellar flight we have discussed the various milestones and deployment strategies as being a long part of the program. It is useful to look at the costs for a staged approach when one possibility is to arrange system level milestones as being related to speed goals (as one example, shown in Figure \ref{fig:mincostvsspeed}). Since spacecraft speed depends on many parameters, such as spacecraft mass and sail material details as well as laser array parameters, the idea of using speed as a milestone needs to be understood more broadly. Nonetheless, it is a useful system  parameter goal as it is simple and tangible. For example, Voyager has achieved approximately 17 km/s or $\beta=5.7\times10^{-5}$. As noted, speed with chemical systems is largely independent of mass and hence useful milestones in speed would be to achieve increasing levels of speed relative to what we have achieved so far. 
    
    For simplicity we will assign a designation of staged system goals for the Starlight program relative to the achieved speed for a putative 1 g wafer, with all the caveats about system speeds being spacecraft mass and sail specific. We assign the system goals as simply:
    \begin{itemize}
        \item Starlight-$x$, where $x=$ percentage of the speed of light achieved for our putative 1 gram wafer.
        \begin{itemize}
            \item For example, a Starlight-1 system would achieve 1\% the speed of light.
        \end{itemize}{}
    \end{itemize}{}    
    Noting the minimum price as above in Equation 57:
    \begin{align}
    \begin{split}{}
        C_T(\$)&=C_1(\textrm{laser})+C_2(\textrm{optics})\\
        &=3C_2(\textrm{optics})=1.5C_1(\textrm{laser})\\
        &=2.70\times10^{11}\beta_0^{4/3}a_1^{2/3}a_2^{1/3}\epsilon_b^{-4/3}\\
        &\hspace{11mm}\times\bigg(\frac{\alpha_d}{\eta}\bigg)^{2/3}\lambda^{2/3}\xi_\textrm{arr}^{1/3}(\xi h\rho m_0)^{1/3},
    \end{split}
    \end{align}
    or for $\epsilon_r=1$ ($\eta=2$), $\alpha_d=1.22$, and $\xi=\xi_\textrm{arr}=\pi/4$ (circular array),
    \begin{align}
    \begin{split}
        \hspace{-1mm}C_T=1.65\times10^{11}\beta_0^{4/3}a_1^{2/3}a_2^{1/3}\epsilon_b^{-4/3}\lambda^{2/3}(h\rho m_0)^{1/3}.
    \end{split}
    \end{align}
    
    In the cost analysis above we had focused on achieving $\beta_0=0.2$ or $0.2c$ which would correspond to our designation of a Starlight-20 system. Noting the cost scaling as 
    \begin{equation}
        C_T\propto\beta_0^{4/3}a_1^{2/3}a_2^{1/3},
    \end{equation}
    we would have a cost (for fixed $a_1$ and $a_2$) for a Starlight-1 being $20^{4/3}\sim54\times$ lower cost than a Starlight-20. In reality, in a staged effort the laser and optics cost metrics will change (lower with time hopefully) both due to technology advancement as well as the economies of scale. While the future is particularly difficult to predict (as opposed to the past) it is useful to look at the system costs for a staged effort. In a real effort, the NRE (human and prototype costs) will be large and our cost analysis will break down. 
    
    A logical approach to a full relativistic mission would be to have precursor missions such as Starlight-0.1 (0.1\%$c$), Star\-light-1 (1\%$c$), etc., as well as use the developed laser array for long range power beaming missions such as powering high $I_\textrm{sp}$ ion engines in the solar system.
    
    {}
    
\section{Energy Per Shot}

    We can calculate the energy per shot which will allow us to calculate the cost per shot and the cost of the energy storage required to store the energy needed per shot. This will allow us to calculate the cost terms $C_3$ and $C_4$ above. Recall we defined $C_3$ as the cost of energy from the ``grid'' including amortization of total number of missions and $C_4$ as the cost of energy stored, including amortization and efficiencies:
    \begin{align}
    \begin{split}{}
        C_3&=a_3P_0t_0\\
        C_4&=a_4P_0t_0.
    \end{split}
    \end{align}
    The total photon energy in the main beam in time $t_0$ is:
    \begin{equation}
        E_\gamma=P_0t_0.
    \end{equation}
    
\subsection{Non-optimized case}

    \begin{equation}
        \beta_0=\bigg(\frac{P_0\eta dD}{c^3\lambda\alpha_d(\xi D^2h\rho+m_0)}\bigg)^{1/2},
    \end{equation}
    \begin{align}
    \begin{split}
        P_0&=v_0^2\frac{c\lambda\alpha_d(\xi D^2h\rho+m_0)}{\eta dD}=\beta_0^2\frac{c^3\lambda\alpha_d(\xi D^2h\rho+m_0)}{\eta dD},
    \end{split}
    \end{align}
    \begin{equation}
        t_0=\frac{v_0}{a}=\bigg(\frac{cdD(\xi D^2h\rho+m_0)}{P_0\eta\lambda\alpha_d}\bigg)^{1/2},
    \end{equation}
    \begin{align}
    \begin{split}
        E_\gamma&=P_0t_0=P_0\bigg(\frac{cdD(\xi D^2h\rho+m_0)}{P_0\eta\lambda\alpha_d}\bigg)^{1/2}\\
        &=P_0^{1/2}\bigg(\frac{cdD(\xi D^2h\rho+m_0)}{\eta\lambda\alpha_d}\bigg)^{1/2}\\
        &=\beta_0\frac{c^2(\xi D^2h\rho+m_0)}{\eta}=\beta_0\frac{mc^2}{\eta}\\
        &=\frac{1}{2}\beta_0mc^2\hspace{2mm}\textrm{for $\epsilon_r=1$}.
    \end{split}
    \end{align}
    Here, $m$ is the total mass ($m_\textrm{sail}+m_0=\xi D^2h\rho+m_0$) and $\eta=2\epsilon_r+(1-\epsilon_r)\alpha$, as before. In the ``roadmap'' we defined the launch efficiency as:
    \begin{align}
    \begin{split}
        \epsilon_\textrm{total}(t_0)&=\frac{\textrm{KE}}{E_\gamma}=\frac{\frac{1}{2}mc^2\beta_0^2}{\beta_0mc^2/\eta}=\frac{1}{2}[2\epsilon_r+(1-\epsilon_r)\alpha]\\
        &=\beta_0\hspace{2mm}\textrm{for $\epsilon_r=1$}.
    \end{split}
    \end{align}
    
\subsection{Optimized case}

    \begin{equation}
        \beta_0=\bigg(\frac{P_0\eta d}{2c^3\lambda\alpha_d}\bigg)^{1/2}(\xi h\rho m_0)^{-1/4},
    \end{equation}
    \begin{align}
    \begin{split}
        P_0&=v_0^2\bigg(\frac{2c\lambda\alpha_d}{\eta d}\bigg)(\xi h\rho m_0)^{1/2}=\beta_0^2\bigg(\frac{2c^3\lambda\alpha_d}{\eta d}\bigg)(\xi h\rho m_0)^{1/2},
    \end{split}
    \end{align}
    \begin{equation}
        t_0=\bigg(\frac{2cd}{P_0\eta\lambda\alpha_d}\bigg)^{1/2}\bigg(\frac{m_0^3}{\xi h\rho}\bigg)^{1/4},
    \end{equation}
    \begin{align}
    \begin{split}
        E_\gamma&=P_0t_0=P_0\bigg(\frac{2cd}{P_0\eta\lambda\alpha_d}\bigg)^{1/2}\bigg(\frac{m_0^3}{\xi h\rho}\bigg)^{1/4}\\
        &=P_0^{1/2}\bigg(\frac{2cd}{\eta\lambda\alpha_d}\bigg)^{1/2}\bigg(\frac{m_0^3}{\xi h\rho}\bigg)^{1/4}=2\beta_0\frac{m_0c^2}{\eta}\\
        &=\beta_0m_0c^2\hspace{2mm}\textrm{for $\epsilon_r=1$}.
    \end{split}
    \end{align}
    Note that since the total system mass in the optimized case is $m=2m_0$, the result is identical to the above non-optimized case.
    \begin{align}
    \begin{split}
        \epsilon_\textrm{total}(t_0)&=\frac{\textrm{KE}}{E_\gamma}=\frac{\frac{1}{2}mc^2\beta_0^2}{2\beta_0m_0c^2/\eta}=\frac{1}{2}[2\epsilon_r+(1-\epsilon_r)\alpha]\beta_0\\
        &=\beta_0 \hspace{2mm}\textrm{for $\epsilon_r=1$}.
    \end{split}
    \end{align}

\subsection{Cost of Energy Used}

    The cost of energy over the life of the system (over all missions) $C_3$ compared to the other system costs will depend on how long the system lasts. If the system lasts indefinitely and is used continuously then energy costs will always dominate, but this is not a fair comparison as the energy costs over the life of the system need to be compared to the energy cost of that same lifetime. The real problem with calculating total energy used ($C_3$) over the lifetime of the system is that this is a dynamic system with exponential growth and hence the basic system is designed to evolve, so there is no real ``final system.'' However, we can make a quick estimate. Assume a 100,000 hour nominal lifetime (a bit optimistic, though some commercial lasers have MTBF in excess of 50,000 hours). Assume the cost of energy is \$0.05/kW-hr or $1.4\times10^{-8}$ \$/J  (commercial cost in low cost (hydro) areas). 100,000 hours of operation would consume \$5/W$_\textrm{elec}$ or \$5 per electrical watt of the laser. At 50\% ``wall plug efficiency'' this would be \$10/W$_\textrm{optical}$ or \$10 per optical watt deployed ($P_\textrm{optical}$). Our goal for the deployed laser cost ($a_1$) is \$0.1/W$_\textrm{optical}$ or \$0.1 per deployed optical watt. Recall:
    \begin{equation}
        C_1=a_1P_\textrm{optical}=\frac{a_1P_0}{\epsilon_b},
    \end{equation}
    where $\epsilon_b$ is the main beam fractional efficiency (fraction of optical power in the main beam where the main beam is the beam on the sail -- power $P_0$ in our notation). This would make the energy cost even more for a desired $P_0$. Hence in the end the energy cost would dominate. However, this not generally how we would ``cost a system,'' but it is still a useful number to understand. 
    
\subsection{Cost of Energy Storage}

    The cost of energy storage (batteries, etc.), $C_4$, is also important to understand. Recall:
    \begin{equation}
        C_4=a_4P_0t_0=a_4E_\gamma,
    \end{equation}
    which includes amortization and efficiencies. We calculated the photon energy $E_\gamma$ per shot above:
    \begin{equation}
        E_\gamma=\beta_0\frac{mc^2}{(2\epsilon_r+(1-\epsilon_r)\alpha)}=\frac{1}{2}\beta_0mc^2\hspace{2mm}\textrm{for $\epsilon_r=1$}.
    \end{equation}
    This applies whether the system is optimized ($m_\textrm{sail}=m_0$) or not. The relationship between mass and speed for a given system is shown in the following:
    
\subsubsection{Non-optimized case}

    \begin{align}
    \begin{split}
        \beta_0=\bigg(\frac{P_0\eta dD}{c^3\lambda\alpha_d(\xi D^2h\rho+m_0)}\bigg)^{1/2},
    \end{split}
    \end{align}
    \begin{align}
    \begin{split}
        E_\gamma=\beta_0\frac{mc^2}{\eta}&=\bigg(\frac{P_0mcdD}{\lambda\alpha_d\eta}\bigg)^{1/2}\sim P_0^{1/2},m^{1/2},d^{1/2}.
    \end{split}
    \end{align}

\subsubsection{Optimized case}

    \begin{equation}
        \beta_0=\bigg(\frac{P_0\eta d}{2c^3\lambda\alpha_d}\bigg)^{1/2}(\xi h\rho m_0)^{-1/4},
    \end{equation}
    \begin{align}
    \begin{split}
        E_\gamma&=2\beta_0\frac{m_0c^2}{\eta}=2\bigg(\frac{P_0\eta d}{2c^3\lambda\alpha_d}\bigg)^{1/2}\frac{m_0c^2}{\eta}(\xi h\rho m_0)^{-1/4}\\
        &=2m_0^{3/4}\bigg(\frac{P_0cd}{2\lambda\alpha_d\eta}\bigg)^{1/2}(\xi h\rho)^{-1/4}\\
        &=\beta_0m_0c^2\hspace{2mm}\textrm{for $\epsilon_r=1$}.
    \end{split}
    \end{align}
    Therefore, for $\epsilon_r=1$:
    \begin{align}
    \begin{split}{}
        E_\gamma&=m_0^{3/4}\bigg(\frac{P_0cd}{\lambda\alpha_d}\bigg)^{1/2}(\xi h\rho)^{-1/4}\sim P_0^{1/2}, m_0^{3/4},d^{1/2},
    \end{split}
    \end{align}
    \begin{equation}
        E_\gamma(\textrm{J})=5.3\times10^{11}m_0^{3/4}\bigg(\frac{P_0d}{\lambda}\bigg)^{1/2}(h\rho)^{-1/4},
    \end{equation}
    for $[m_0]=$ g, $[P_0]=$ GW, $[d]=$ km, $[\lambda]=[h]=$ $\mu$m, and $[\rho]=$ g/cc.
    
    Note that the energy storage needed and thus cost for energy storage is proportional to $P_0^{1/2}$, $d^{1/2}$, and $m_0^{3/4}$ (optimized case) and $(mD)^{1/2}$ (non-optimized), while laser cost is proportional to the power $P_0$ and optics cost is proportional to optics area, or $d^2$. Thus, energy storage costs scale much more slowly than laser and optics costs. However, energy storage costs depend on the mission mass for a given system ($P_0$, $d$). Hence the storage costs are only fixed if the mission mass is fixed (once the system is built). Current battery storage costs (Li in large quantities) are less than $C_4=\$0.1$/W-hr $\sim2.8\times10^{-5}$ \$/J.
    
    Recall $E_\gamma$ is the main beam photon energy to $L_0$ (also $\beta_0$, $t_0$). With the efficiencies of the main beam fraction as well as the laser, power supply, and charge efficiencies, the actual energy storage needs will be greater (perhaps a factor of  2-3 higher). We can now calculate the energy storage costs (for $\epsilon_r=1$):
    
    \begin{align}
    \begin{split}
        \hspace{-1mm}E_\textrm{storage}(\textrm{J})&=\frac{E_\gamma}{\epsilon_\textrm{storage}}\\
        &=\frac{5.3\times10^{11}}{\epsilon_\textrm{storage}}m_0^{3/4}\bigg(\frac{P_0d}{\lambda}\bigg)^{1/2}(h\rho)^{-1/4},
    \end{split}
    \end{align}
    where $\epsilon_\textrm{storage}$ includes all relevant efficiencies. Assuming storage costs of \$0.1/W-hr ($2.8\times10^{-5}$ \$/J):
    \begin{align}
    \begin{split}
        \hspace{-2.75mm}E_\textrm{storage}(\$)&=a_4E_\gamma(\textrm{J})\\
        &=5.3\times10^{11}a_4m_0^{3/4}\bigg(\frac{P_0d}{\lambda}\bigg)^{1/2}(h\rho)^{-1/4}\\
        &\sim\frac{1.5\times10^7}{\epsilon_\textrm{storage}}m_0^{3/4}\bigg(\frac{P_0d}{\lambda}\bigg)^{1/2}(h\rho)^{-1/4},
    \end{split}
    \end{align}
    for $a_4=2.8\times10^{-5}$ \$/J (\$0.1/W-hr). We can also write this for a given $\beta_0$ and $m_0$ as:
    \begin{align}
    \begin{split}
        E_\textrm{storage}(\$)&=a_4E_\gamma(\textrm{J})=a_42\beta_0\frac{m_0c^2}{\eta}\\
        &=a_4\beta_0m_0c^2\hspace{2mm}\textrm{for $\epsilon_r=1$}.
    \end{split}
    \end{align}
    
\begin{figure}
        \centering
        \includegraphics[width=0.45\textwidth]{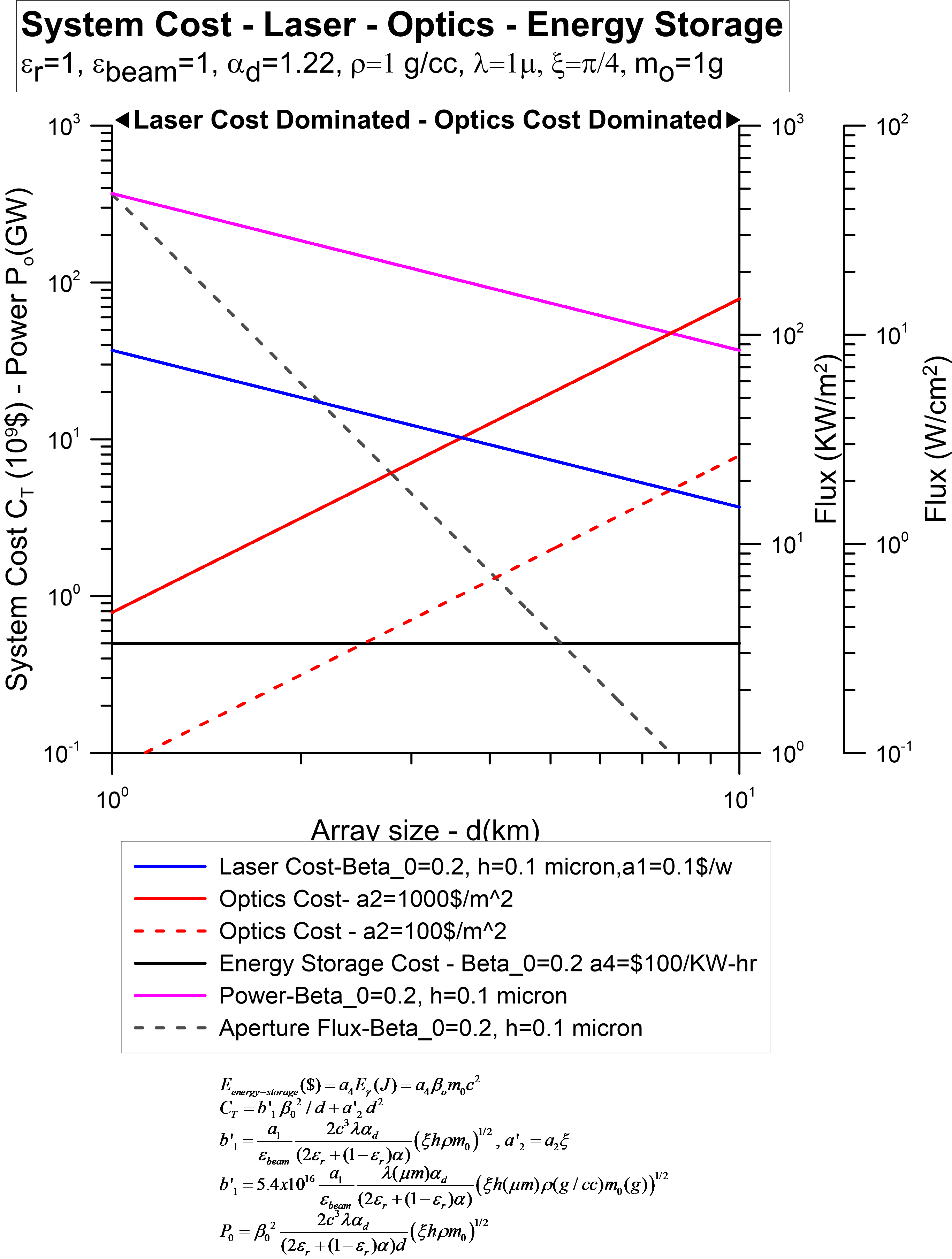}
        \caption{Laser cost (blue), optics cost (red), and energy storage cost (black) as a function of array size in km. Also shown is the equivalent array power (magenta) and aperture flux (dashed black).}
        \label{fig:costwithenergystorage}
    \end{figure}

    Thus, even a 100 GW, 10 km array for driving 1 g payloads would only require energy storage costs that are roughly \$0.5B. This justifies our assumption above that the energy storage costs are sub-dominant in the overall system cost to the laser and optics costs for small mass payloads. This is shown clearly in the figure comparing laser, optics, and energy storage costs. The cross over between laser and optics costs is also clear. The effect of lower optics costs in lowering both the laser costs (large sized array = less laser power needed = less laser cost) and the total system cost is also clear. The system cost minimum is NOT at the intersection between the laser and optics costs as the minimum also depends on the cost slopes. The cost minimum is for laser cost = 2/3 and optics costs = 1/3 of the total cost for the case where the energy storage costs are a small fraction of the total cost (as is the case shown in Figure \ref{fig:costwithenergystorage}). See discussion on cost minimization and the figure above.
    
    {}
    
\subsection{Energy Storage for Large Mass Payloads}

    Since the acceleration time $t_0$ (and hence the energy storage required and the energy storage costs) scale as $m_0^{3/4}$, large mass payloads (such as kg or larger) require much longer illumination times. Hence, IF we wanted to store the energy required for such launches then the costs would dramatically escalate. For example, a $m_0=1$ kg payload would cost about 180 times more than the energy storage for a 1 gram payload. It would also require 180 times longer illumination time. The energy storage cost ($\sim$ \$100B) would then dominate the system costs. However, such systems do not make much sense for ground based systems due to the long exposure times needed given the large rotation rate of the Earth. A polar deployed system (South Pole or nearby) would be possible, though as discussed it has large deployment challenges. A lunar based system would make sense (except for the deployment costs) for launching large mass payloads. There is a cross over point where power production (power plant) costs are more cost effective than energy storage facilities. This point is specific to the type of power plant (solar, fossil fuel, nuclear) as well as the plant size and storage technology.
    
\subsection{Storage Amortization and Energy Production Costs}

    When the storage costs are a small fraction of the system cost, as they are for low mass launches, storage is reasonable, but once the costs become a significant or dominant cost driver the many issues associated with storage (in particular finite cycle life with batteries) need to be carefully weighed against the cost of power production. Power production will be critically dependent on the site location (Earth, orbital, lunar). One good cost metric is the current push to bring the cost of terrestrial solar installations to below \$1/W$_\textrm{installed}$. For instance, in the example above where the storage costs for relativistic 1 gram payloads were around \$1B, the amount of solar that could be installed is small compared to the system needs, but when launching larger payloads (kg level) the storage costs could be \$100B and the amount of solar that could be installed for the same price becomes comparable to the power requirements of the full system. For orbital or lunar deployment the situation is vastly more complex with the deployment costs dwarfing the system costs. 
    
\subsection{Cost Analysis Including Energy Used and Energy Storage}

    \begin{align}
    \begin{split}
        C_T&=C_1+C_2+C_3+C_4\\
        &=\frac{a_1P_0}{\epsilon_b}+a_2\xi_\textrm{arr}d^2+a_3P_0t_0+a_4P_0t_0\\
        &=\frac{a_1P_0}{\epsilon_b}+a_2\xi_\textrm{arr}d^2+(a_3+a_4)E_\gamma,
    \end{split}
    \end{align}
    where
    \begin{align}
    \begin{split}
         E_\gamma&=P_0t_0=\beta_0\frac{mc^2}{(2\epsilon_r+(1-\epsilon_r)\alpha)}=\frac{1}{2}\beta_0mc^2\hspace{2mm}\textrm{for $\epsilon_r=1$}.
    \end{split}
    \end{align}
    Notice that we can combine the amortized energy used cost term $a_3$ and the stored energy cost term $a_4$. For the optimized case:
    \begin{equation}
        \beta_0=\bigg(\frac{P_0\eta d}{2c^3\lambda\alpha_d}\bigg)^{1/2}(\xi h\rho m_0)^{-1/4},
    \end{equation}
    \begin{align}
    \begin{split}
        P_0&=\beta_0^2\bigg(\frac{2c^3\lambda\alpha_d}{\eta d}\bigg)(\xi h\rho m_0)^{1/2},
    \end{split}
    \end{align}
    \begin{align}
    \begin{split}
         E_\gamma&=P_0t_0=2\beta_0\frac{m_0c^2}{(2\epsilon_r+(1-\epsilon_r)\alpha)}=\beta_0m_0c^2\hspace{2mm}\textrm{for $\epsilon_r=1$}.
    \end{split}
    \end{align}
    \begin{align}
    \begin{split}
        C_T&=\frac{a_1P_0}{\epsilon_b}+a_2\xi_\textrm{arr}d^2+(a_3+a_4)E_\gamma\\
        &=\frac{a_1}{\epsilon_b}\beta_0^2\bigg(\frac{2c^3\lambda\alpha_d}{\eta d}\bigg)(\xi h\rho m_0)^{1/2}+a_2\xi_\textrm{arr}d^2\\
        &\hspace{5mm}+(a_3+a_4)2\beta_0\frac{m_0c^2}{\eta}.
    \end{split}
    \end{align}
    Note that the system cost including energy used and storage is simply the system cost $C_1+C_2$ (no energy use nor energy storage cost) plus the cost of energy use and energy storage, $C_3+C_4$.
    
    This means the array size $d$ where the cost $C_T$ is minimized is the same with or without including the cost of energy use and storage since the cost of energy use and storage is simply an addition (for fixed $\beta_0$, $m_0$) to the system cost without energy use and storage since there is no dependence of the energy use and storage cost on array size $d$. While the total cost increases, the array size $d$ and hence the power at the cost minimum are still the same. Note the above is for a single shot. For $N_\textrm{shot}$ number of shots, the total cost is:
    \begin{equation}
        C_T=\frac{a_1P_0}{\epsilon_b}+a_2\xi_\textrm{arr}d^2+(N_\textrm{shot}a_3+a_4)E_\gamma.
    \end{equation}
    This has the same dependence as one shot and thus the cost minimum is still same, but the total cost is larger by the total cost of energy used. \textbf{The cost minimum (with or without including the cost of energy use and storage) is still when the laser cost is twice the cost of the optics.}
    
\section{Cost Comparison to Recent and Past NASA Programs}

    Any realistic directed energy propulsion system to reach relativistic speeds will be expensive. It is useful to compare to some of the larger NASA programs to some of the system cost estimates. A critical difference is that the cost of the R\&D phase of the DE program we are discussing, that will likely last over several decades, will be coupled to an exponential technology which is unlike any past NASA program. The DE will also be driven by other market factors (telecom, high speed photonic interfaces in commercial electronics, etc.) and other forces will push the ``DE market'' including the DoD sectors. Any realistic cost estimate for such a long term program is not credible currently. Our numbers are based on both realistic cost expectations from existing DE and optics technologies as well as the assumed exponential growth in photonics. The latter, as discussed, is already going on and will almost certainly continue for reasons that have nothing to do with DE driven propulsion. For example, the current worldwide photonics market is nearing \$180B/yr (2016 USD) with Si photonics expected to exceed \$1T by 2022 with annual growth rates of approximately 20\%. \textbf{These numbers dwarf the current entire chemical launch industry and show the ``engine'' upon which a DE program would be propelled economically.}
    
    For historical reference, we note the Apollo program, built upon a very large and expensive infrastructure that was primarily DoD in its origins (ICBM's),  cost (NASA side only) about \$200B in 2016 USD. The Shuttle program cost about \$210B in 2010 USD. US cost of the ISS was about \$75M in 2014 USD. The JWST will cost in total close to \$10B in the end.
    
    Another critical factor is if a DE launch were to be space based (vs ground based) then the launch and space deployment costs will likely completely dominate the program. Ground based DE, as discussed, is only feasible IF we can overcome the atmospheric perturbations and then will only support low mass/high speed systems due to the short illumination times needed (Earth rotation issues). Otherwise, a ground-based system could be used for hybrid propulsion (mass ejection) high mass launches, as previously discussed.

\section{\label{sec:conclusion}Conclusion}

Directed energy enables a path forward in a variety of new technologies and mission spaces, including long-range power beaming, laser communications, rapid interplanetary transit, m\-anipulation and composition analysis of asteroids and comets, planetary defense, and interstellar flight. The combination of building upon a modular and scalable design philosophy and the additional dependence on photonics, which is an exponentially expanding area driven by vast consumer demand, allows for incredible scalability of near-future systems. We have shown how integrated photonics and mass production will be fundamentally necessary to afford the full-scale realization of this vision and we have derived an analytical cost model which is driven by the fundamental physics of the proposed system. This allows us to make economically informed decisions and create a logical path forward to interstellar flight.

\section{Acknowledgments}

Funding for this program comes from NASA grants NIAC Phase I DEEP-IN – 2015 NNX15AL91G and NASA NIAC Phase II DEIS – 2016 NNX16AL32G and the NASA California Space Grant NASA NNX10AT93H and from the Emmett and Gladys W. Technology Fund and Breakthrough Initiatives as well as the Limitless Space Institute.

\appendix

 \bibliographystyle{elsarticle-num-names} 


\bibliography{economicsreferences}




\end{document}